\documentclass[11pt,a4paper]{article}

\usepackage[utf8]{inputenc}
\usepackage{geometry}
\usepackage{booktabs}
\usepackage{amsmath}
\usepackage{amssymb}
\usepackage{setspace}
\usepackage{graphicx}
\usepackage{lineno}
\usepackage{makecell}
\usepackage{ulem}
\usepackage{xcolor}
\usepackage{cancel}
\usepackage{bm}
\usepackage{mathrsfs}

\usepackage{pdflscape}
\usepackage[ruled,vlined]{algorithm2e}
\usepackage{multirow}
\usepackage{subcaption}
\usepackage{hyperref}

\title{Multi-outcome trials with a generalised number of efficacious outcomes}
\author{Martin Law, Michael J. Grayling, Adrian P. Mander}
\date{December 2020}

\begin{document}

\maketitle

\begin{abstract}
Existing multi-outcome designs focus almost entirely on evaluating whether all outcomes show evidence of efficacy or whether at least one outcome shows evidence of efficacy. While a small number of authors have provided multi-outcome designs that evaluate when a general number of outcomes show promise, these designs have been single-stage in nature only. We therefore propose two designs, of group-sequential and drop the loser form, that provide this design characteristic in a multi-stage setting. Previous such multi-outcome multi-stage designs have allowed only for a maximum of two outcomes; our designs thus also extend previous related proposals by permitting any number of outcomes.
\end{abstract}

\section{Introduction}
In this manuscript we will consider single-arm trials where there are multiple key outcomes to account for. This is important as a clinical trial will typically measure many outcomes of interest. This may be done for a number of reasons. For example, investigators may plan to conduct a phase III trial using the outcome that will show the strongest evidence of treatment efficacy, and a multiple outcome trial in phase II will help identify this candidate outcome. Alternatively, investigators may wish to measure multiple outcomes in a phase III trial with the intention of declaring trial success where a promising treatment effect is observed on one of the outcomes. Furthermore, some disease conditions are typically assessed in a multi-dimensional manner, for example respiratory health~\cite{Celli932}. There may also exist a core outcome set for the condition of interest, detailing a set of outcomes that should be measured when evaluating a treatment for that condition~\cite{Williamson2017}.
In general, there may simply be interest in observing a novel treatment's effects on a range of relevant endpoints.

\subsection{Existing designs}
\subsubsection{Multi-outcome single-stage trials}
A trial with multiple outcomes may be designed to evaluate whether a positive effect is observed on at least one of several key outcomes. Outcomes of this type are known as ``multiple primary" outcomes~\cite{Sozu}. The presence of multiple primary outcomes means that additional testing must be accounted for compared to trials with a single outcome. Specifically, in the case of multiple primary outcomes, we must consider the family-wise error-rate, the probability of at least one type-I error occurring. For example, to control this as desired, we might apply a multiple testing correction such as the Bonferroni procedure.

In contrast to multiple primary outcomes, a multiple outcome trial may be designed to evaluate whether a positive effect is observed for all outcomes in some specified set. In this case, the outcomes are described as ``co-primary"~\cite{Sozu}. Sozu et al. have examined multiple outcome trials in detail, providing design approaches for both multiple and co-primary outcome trials~\cite{Sozu}.

\subsubsection{Group sequential trials}
Group sequential trials, also known as multi-stage trials, include multiple interim analyses. Such trials may permit early stopping at these interim analyses if the current estimate of the treatment effect is greater than some upper boundary only, less than some lower boundary only, or either, where such boundaries are specified in advance~\cite{jennturn}. Group sequential trials improve upon single-stage trials by permitting such early stopping at each interim analysis; typically this means that the expected (or average) sample size required by a multi-stage trial is below that required by the corresponding single-stage trial.

\subsubsection{Multi-arm drop the loser trials}
If more than one experimental treatment exists for a given condition of interest, then an improvement on undertaking a series of single-arm or two-arm trials is to use a multi-arm design. Multi-arm trials allow multiple experimental treatments to be simultaneously compared to a common control treatment, reducing the required sample size compared to conducting a series of trials each with a single experimental treatment arm. The concepts of multi-arm and multi-stage trials can be combined to create a trial design containing multiple experimental treatment arms, tested over multiple interim analyses, or stages. Such trials are known as multi-arm multi-stage or MAMS trials. 

In MAMS trials, where multiple experiment treatment arms are evaluated simultaneously, the number of arms may be reduced over the duration of the trial. This is typically undertaken by ceasing recruitment on an arm or arms, based on current results. This is known as \textit{dropping} an arm (or arms). This provides a statistical advantage, as more participants can subsequently be recruited to the remaining, better-performing arms, providing more information about those arms. A disadvantage to dropping arms in general is that the required sample size is not fixed: a trial consisting of mostly well-performing treatments may have little or no dropping of arms. Conversely, a trial consisting of mostly poorly-performing treatments may drop many arms in the early stages. This makes the certain practical aspects of the design, such as trial duration and cost, uncertain.

One approach to dropping treatment arms is the \textit{drop the loser} (DtL) design, where exactly one treatment arm is dropped at each stage or the number of arms at each stage is otherwise pre-determined~\cite{Wason2017loser}. As a result, the number of stages required for the best-performing treatment (or treatments) to reach some required number of participants is fixed and known in advance, even if the identity of that treatment (or those treatments) is not known in advance. This aspect of the DtL design allows DtL MAMS trials to be planned with more certainty than other MAMS designs that do not have this property.

\subsubsection{Dropping outcomes}\label{sec:dropping_outcomes}
Multi-outcome trials may continue to measure all planned outcomes even when some of the outcomes have a low probability of contributing to trial success, either in the multiple primary or multiple co-primary outcome setting. If some outcomes are particularly expensive, invasive or time-consuming to measure, it may be valuable to use a trial design that stops measuring outcomes that are performing poorly. We describe this action as \textit{dropping an outcome}, similar to the dropping of arms in the multi-arm setting.

\subsubsection{Multi-outcome multi-stage trials}

The approach of Sozu et al. to multiple primary and co-primary outcomes in clinical trials focuses on single-stage, two-arm designs, and they describe this work as a foundation for other design features, including group sequential trials~\cite{Sozu}. A review by Hamasaki et al. of clinical trial designs that use co-primary endpoints describes numerous approaches to multi-outcome design, including multi-stage designs~\cite{Hamasaki2017}. Among these are designs that include early stopping for a go decision only or for either a go or no go decision, for either two outcomes or two or more outcomes. These designs are classified in Table~\ref{tab:hamasaki_summary}. In all cases, the designs use co-primary endpoints, and thus promising results must be observed on every endpoint for the null hypothesis to be rejected. There is no framework to test if some subset of outcomes show promising effects.

\begin{table}[htbp!]
    \centering
    \begin{tabular}{l|lr}
    \toprule
        Author & Early stopping permitted & Number of outcomes\\
        \midrule
        Ando et al.\cite{ando2010practical,ando2015sample} & Go decision only & 2\\
        Asakura et al.\cite{asakura2014sample} & Go decision only & 2\\
        Cheng et al.\cite{cheng2014statistical}   & Go decision only & 2\\
        Hamasaki et al.\cite{hamasaki2015group}& Go decision only & $\geq2$\\
        Cook and Farewell~\cite{cook1994guidelines}& Go or no go decision & 2\\
        Jennison and Turnbull~\cite{jennison1993group}& Go or no go decision & 2\\
        Schuler et al~\cite{schuler2017choice}& Go or no go decision & 2\\
        Asakura et al.\cite{asakura2017interim} & Go or no go decision & $\geq2$\\
    \bottomrule
    \end{tabular}
    \caption{Summary of group sequential designs for co-primary outcomes by Hamasaki et al.~\cite{Hamasaki2017}}
    \label{tab:hamasaki_summary}
\end{table}

\subsubsection{Separate vs. simultaneous stopping}
Outside the case of multiple primary endpoints, where only a single outcome must show promise for the trial to be a success, a decision must be made regarding when to conclude that the necessary number of outcomes show promise. It is possible to conclude that an outcome is promising as soon as its test statistic is found to exceed a corresponding efficacy stopping boundary. The outcome may cease to be measured at this point, which we describe as dropping an outcome, as described above (Section~\ref{sec:dropping_outcomes}). Conversely, one may choose not to make conclusions regarding every outcome separately, but instead deem the trial a success only if enough outcome test statistics exceed their corresponding efficacy stopping boundaries simultaneously. In the multi-outcome case, this choice has been discussed previously~\cite{asakura2014sample,hamasaki2015group,hamasaki2016group}. This can be viewed as related to the two options for stopping MAMS trials, known as \textit{separate} versus \textit{simultaneous} stopping~\cite{urach2016multi,wason2016some}. In the area of MAMS, this choice is generalised by Grayling et al.~\cite{grayling2017efficient}, to permit stopping after a specified number of arm-specific null hypotheses have been rejected.

\subsubsection{Number of outcomes required to show promise for trial success}
The review of multi-outcome designs by Hamasaki et al.~\cite{Hamasaki2017} covers co-primary endpoints only, where the trial is a success only if a certain degree of efficacy is shown on all measured outcomes. Sozu et al.~\cite{Sozu} describe designs for co-primary endpoints and for multiple primary endpoints, where trial success is declared if a promising size is found for any measured outcome. Historically, the focus of multi-outcome design is on co-primary endpoints and multiple primary endpoints. In contrast, Delorme et al.~\cite{delorme2016type} describe a generalised power approach, a multi-outcome design where trial success is declared if some specified number of outcomes (or more), out of a larger set of outcomes show promise. The approaches we propose in this work are centred on this idea.

\subsubsection{Multi-arm multi-stage trials with generalised error-rates}
In addition to Delorme et al.'s generalised power approach~\cite{delorme2016type}, Grayling et al.~\cite{grayling2017efficient} present an approach to MAMS (rather than multi-outcome single-stage) trials that allows trials to be powered to find any specified number of efficacious arms. This design also features a generalised approach to stopping, permitting stopping as soon as any specified number of separate hypotheses are rejected.


\subsubsection{Composite outcome trials}
When multiple outcomes are to be measured, a choice must be made regarding whether or not to combine the measurements into a single composite outcome. Testing a single, composite measurement is statistically simpler, and may be appropriate when the outcomes are deemed suitable to combine~\cite{jennturn}. However, both determining how to appropriately combine outcome measurements and interpreting the resulting composite outcome may still be difficult. Consequently, the multiple-outcome design may be preferred in the case that creating a composite outcome is either inappropriate or otherwise challenging. Composite designs may have multiple stages, and we will examine such a design.


\subsection{Brief description of existing multi-outcome multi-stage designs}
The main limitation of existing work is that current multi-outcome multi-stage designs focus almost entirely on evaluating if all outcomes show evidence of efficacy or if at least one outcome shows evidence of efficacy. While Delorme et al.~\cite{delorme2016type} provide a multi-outcome design that evaluates when a general number of outcomes show promise, this design is single-stage only. Using a single-stage design means that there are no interim analyses and no decisions made until the end of trial. In single-stage multi-outcome trials, the sample size is fixed and every outcome is measured for every participant. We propose two designs that provide this design characteristic in a multi-stage setting. Beyond this, many multi-outcome multi-stage designs allow only a maximum of two outcomes, while the proposed designs permit any number of outcomes. Finally, one of the two proposed designs permits ceasing measurement of an outcome that is performing poorly. While this design characteristic is not novel on its own, we believe that this property has not been implemented in a design that evaluates multiple outcomes powered under the condition of a general number of outcomes showing promise.

\subsection{Proposed designs}
In both proposed designs, we subsume the concepts of co-primary and multiple primary outcomes into a general framework of single-arm designs that permit rejection of a null hypothesis $H_0$ when promising effects are observed on some specified $m$ out of $K$ outcomes. We apply this concept to multi-outcome multi-stage design, allowing the trial to end at any stage, for either a go decision (reject $H_0$) or a no go decision (do not reject $H_0$). The first proposed design permits any number of stages $J$. This design will be compared to a multi-stage composite design, where again the trial may end at any stage, for a go or no go decision, and a single, composite outcome is evaluated at each stage. The second proposed design limits the number of stages to two, and permits dropping poorly-performing outcomes at the interim analysis while still allowing the trial to end at this point for a go decision or no go decision. This design is compared to a multi-outcome single-stage design that, like both proposed designs, rejects the null hypothesis when promising effects are observed on $m$ out of $K$ outcomes.

\section{Methods: Multi-outcome multi-stage design with general number of required efficacious outcomes}\label{sec:MOMS_design_1}

Let $K$ be the total number of (continuous) outcomes that will be measured in the trial. Let $J$ be the maximum number of allowed stages of the design. The number of participants in each stage of the trial is denoted by $n$. The maximum sample size is then $N=Jn$. We let $X_{ik}, i=1, \dots, Jn, k=1, \dots, K$ be the response in participant $i$ for outcome $k$. The responses are assumed to have the following multivariate normal distribution:

\begin{align*}
\begin{pmatrix}
X_{i1}\\
X_{i2}\\
\vdots\\
X_{iK}
\end{pmatrix}
 &\sim  MVN_K
\begin{bmatrix}
\begin{pmatrix}
\mu_1\\
\mu_2\\
\vdots\\
\mu_K
\end{pmatrix}\!\!,&
\begin{pmatrix}
\sigma_1^2 	& \rho_{12}\sigma_1\sigma_2 & \hdots & \rho_{1K}\sigma_1\sigma_K\\
\rho_{21}\sigma_2\sigma_1 	& \sigma_2^2 & \hdots & \rho_{2K}\sigma_2\sigma_K\\
\vdots	& \vdots & \ddots & \vdots\\
\rho_{K1}\sigma_K\sigma_1 & \rho_{K2}\sigma_K\sigma_2 & \hdots & \sigma_K^2
\end{pmatrix}
\end{bmatrix} . \\[2\jot]
&
\end{align*}

As noted, we assume that interest lies in whether $m$ or more outcomes show promise. Using a single hypothesis approach, the null and alternative hypotheses are

\begin{equation}
H_0: \sum^K_{k=1} \mathbb{I}(\mu_k > 0) < m, \qquad \qquad H_1: \sum^K_{k=1} \mathbb{I}(\mu_k > 0) \geq m. \\
\label{eq:H0_moms}
\end{equation}

After each stage $j$, an interim analysis is undertaken at which point the trial may stop for either a go decision or no go decision. The lower and upper stopping boundaries at stage $j$ are denoted $f_j$ and $e_j$ respectively. The test statistic for outcome $k$ at stage $j$ is $Z_{jk} = \hat{\tau}_{jk} \sqrt{\mathcal{I}_j} = \hat{\tau}_{jk} \sqrt{jn/\sigma_k}$, where $\hat{\tau}_{jk}=\sum_{i=1}^{jn} x_{ik}/jn$ is the observed effect for outcome $k$ at analysis $j$.  The trial will end and the null hypothesis will be rejected if $m$ of the test statistics $Z_{jk}$ simultaneously exceed upper stopping boundary $e_j$, i.e. if

\begin{equation*}
\sum^K_{k=1} \mathbb{I}(Z_{jk} > e_j) \geq m, \text{for any } j.
\end{equation*}

Conversely, a trial will end and the null hypothesis will not be rejected if $K-m+1$ outcomes are simultaneously lower than lower stopping boundary $f_j$, i.e. if
 
 \begin{equation*}
\sum^K_{k=1} \mathbb{I}(Z_{jk} < f_j) \geq K-m+1, \text{for any } j.
\end{equation*}

\begin{figure}[htbp!]
    \centering
    \begin{subfigure}[htbp!]{0.49\textwidth}
    \includegraphics[width=\textwidth]{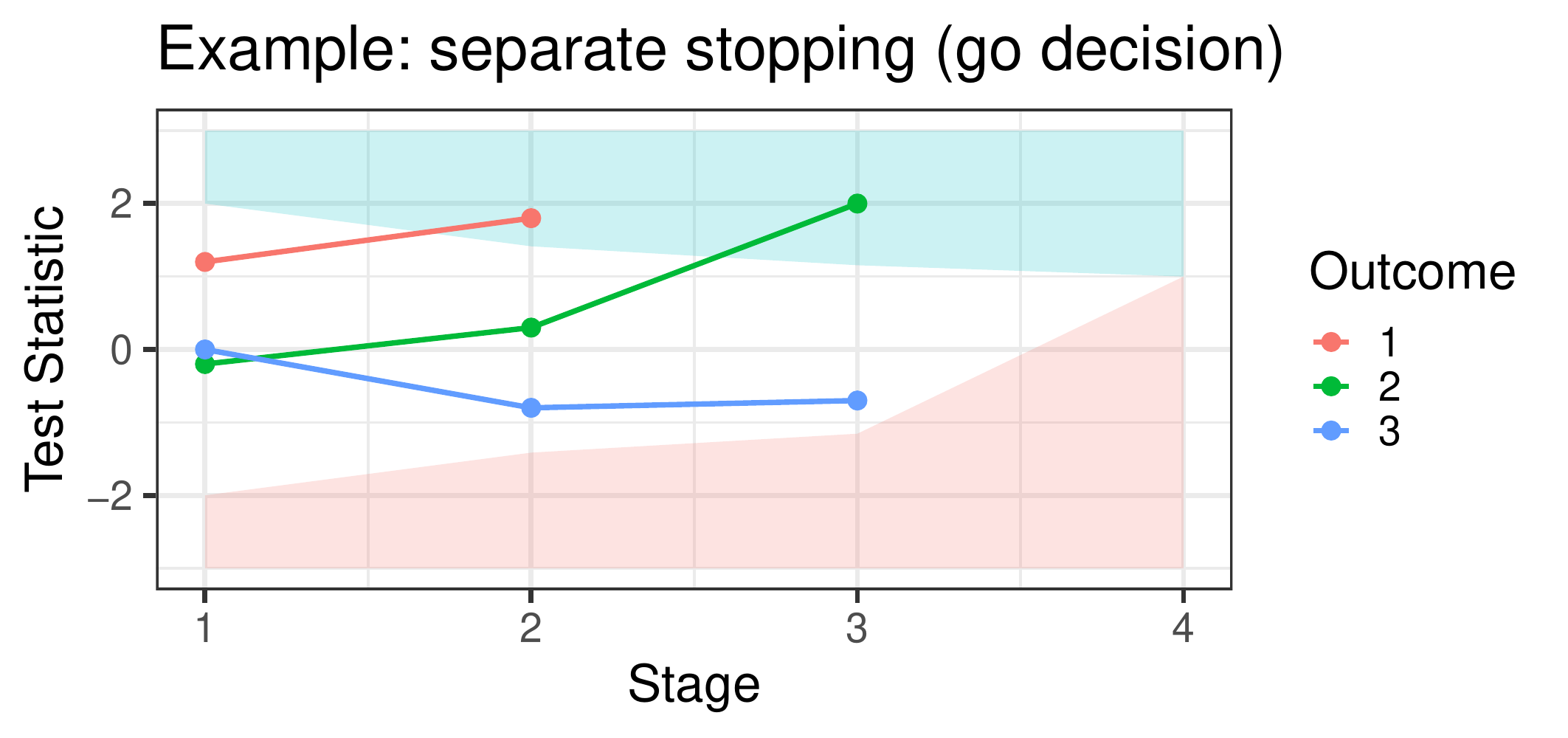}
    \caption{Example of separate stopping approach. Go decision at stage 3.}
    \label{fig:separate_stopping}
    \end{subfigure}
    ~
    \begin{subfigure}[htbp!]{0.49\textwidth}
    \includegraphics[width=\textwidth]{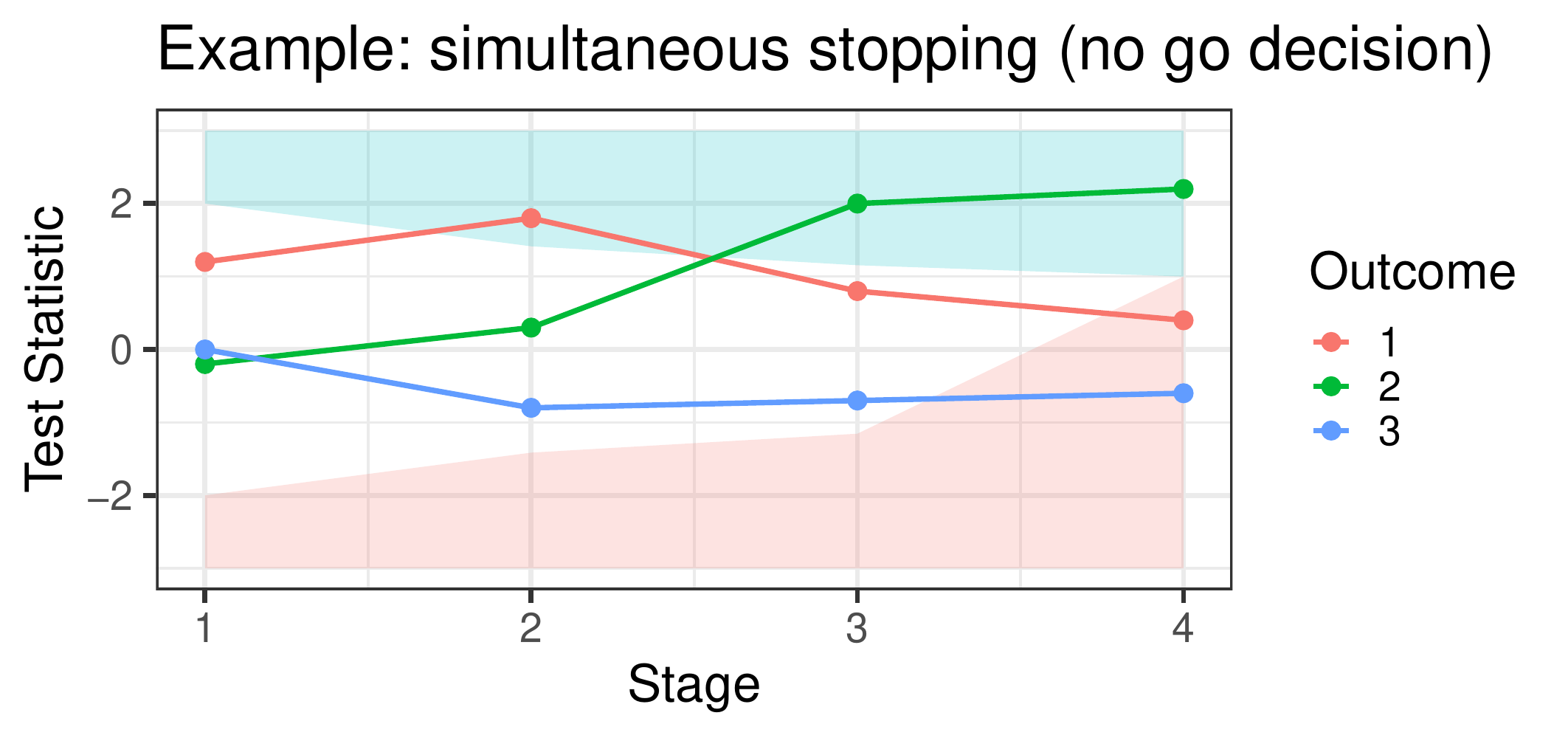}
    \caption{Example of simultaneous stopping approach. No go decision at stage 4.}
    \label{fig:simultaneous_stopping_nogo}
    \end{subfigure}  
    ~
    \begin{subfigure}[htbp!]{0.49\textwidth}
    \includegraphics[width=\textwidth]{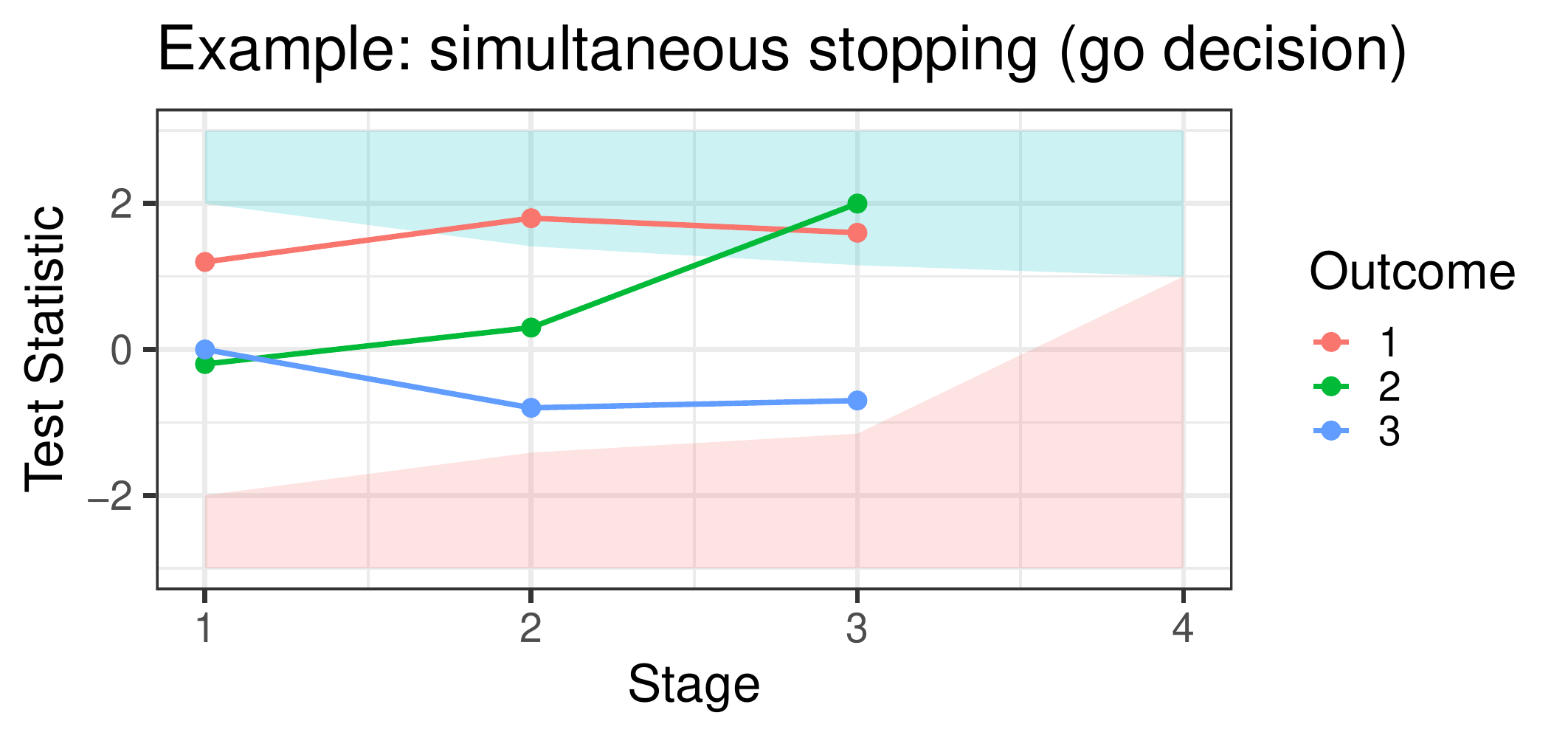}
    \caption{Example of simultaneous stopping approach. Go decision at stage 3.}
    \label{fig:simultaneous_stopping_go}
    \end{subfigure}
    \caption{Examples of separate and simultaneous stopping approaches.}
    \label{fig:compare_separate_simultaneous}
\end{figure}

This is a \textit{simultaneous} multi-stage approach. This is in contrast to a \textit{separate} multi-stage approach, where there are $K$ separate hypotheses, one for each outcome, each of which may be rejected (or not) independently of one another~\cite{hamasaki2015group,hamasaki2016group}. In a separate approach, a decision to reject or not reject some hypothesis $H_k, \, k=1, \dots, K$ is permitted at any stage, and occurs when the corresponding test statistic crosses an upper or lower stopping boundary. Once a decision has been made regarding $H_k$, measurement of outcome $k$ will end. This reduces the number of outcome measurements made. Reducing the expected number of measurements (ENM) may be of particular interest if there are some outcomes that we desire to minimise, either due to cost or otherwise~\cite{hamasaki2018review}, and using the separate multi-stage approach is one way of doing this. A visual comparison is provided in Figure~\ref{fig:compare_separate_simultaneous}, where we show an example design with $J=4$ stages, $K=3$ outcomes and number of outcomes required to show promise $m=2$. In this example, and for the first proposed design, we use stopping boundaries of the form proposed by Wang and Tsiatis~\cite{Wang1987}, for which the stopping boundaries can be characterised by scalars $C$ and $\Delta$: $e_j=Cj^{\Delta-0.5}$, $j=1, \dots , J$. Similarly $f_j=-Cj^{\Delta-0.5}$ for $j=1, \dots , J-1$ and $f_j=e_J$ for $j=J$ to ensure a decision is reached by the final stage. This is a generalisation of the boundaries proposed by Pocock~\cite{pocock1977group} and by O'Brien and Fleming~\cite{obrien1979multiple}, which are special cases equivalent to $\Delta=0.5$ and $\Delta=0$ respectively. Outcome-specific boundaries $C_k, \, k=1, \dots, K$, could theoretically be obtained through $K$-dimensional optimisation. However, given the definition of type-I error-rate used, detailed in Section~\ref{sec:MO_typeIerror_power}, there are potentially infinite sets of $K$ constants that satisfy any required type-I error-rate. Therefore, we do not consider this possibility further. Figure~\ref{fig:separate_stopping} shows the separate approach: at stage 2, outcome 1 crosses the upper boundary and is no longer measured; at stage 3, a second outcome cross the upper boundary, meaning that $m=2$ outcomes have separately shown promise, and the trial ends. In Figures~\ref{fig:simultaneous_stopping_nogo} and~\ref{fig:simultaneous_stopping_go}, the same initial data are shown, but in a simultaneous approach. Now, having a single outcome cross the upper boundary at stage 2 (or any stage) has no effect on the subsequent number of outcomes measured -- all outcomes continue to be measured until either $m$ outcomes simultaneously cross the upper boundary or $K-m+1=2$ outcomes cross the lower boundary. In the example in Figure~\ref{fig:simultaneous_stopping_nogo}, outcome 1 crosses back over the upper boundary at stage 3, and the trial continues. At stage 4, $K-m+1$ outcomes simultaneously cross the lower boundary and consequently a no go decision is made. In the example in Figure~\ref{fig:simultaneous_stopping_go}, outcome 1 remains above the upper boundary at stage 3, and so $m$ outcomes have simultaneously crossed the upper boundary, and consequently a go decision is made.

\subsection{Covariance structure}
The covariance structure must be derived for the multivariate normal distribution of the test statistics across differing stages and outcomes. Although the proposed designs have sample size $jn$ at stage $j$, that is, with an equal number of participants at each stage, for the sake of generality we derive the covariance matrix for a general number of participants at each stage. Define $n_j, \, j=1, \dots, J$ to be the sample size at stage $j$, and $N_j$ to be the total sample size at stage $j$, that is, $N_j = n_1 + n_2 + \dots + n_j = \sum_{i=1}^j n_i$. For a single outcome $k$, the covariance of two test statistics at stages $j_A, j_B, \, j_B \geq j_A$ is 

\begin{align}\label{eq:one_outcome}
    \text{cov}(Z_{j_Ak}, Z_{j_Bk}) & = \text{ cov} \left( \sqrt{\frac{N_{j_A}}{\sigma^2_k}} \hat{\mu}_{j_A k},  \sqrt{\frac{N_{j_B}}{\sigma^2_k}} \hat{\mu}_{j_B k} \right) \nonumber \\
    & = \sqrt{\frac{N_{j_A}}{\sigma^2_k}} \sqrt{\frac{N_{j_B}}{\sigma^2_k}} \text{ cov} \left( \hat{\mu}_{j_A k} ,  \hat{\mu}_{j_B k}  \right) \nonumber \\
    & = \sqrt{\frac{N_{j_A}}{\sigma^2_k}} \sqrt{\frac{N_{j_B}}{\sigma^2_k}} \text{ cov} \left(\frac{1}{N_{j_A}} \sum_{i=1}^{N_{j_A}} X_{ik},  \frac{1}{N_{j_B}} \sum_{i=1}^{N_{j_B}} X_{ik} \right) \nonumber \\
    & = \sqrt{\frac{N_{j_A}}{\sigma^2_k}} \sqrt{\frac{N_{j_B}}{\sigma^2_k}} \frac{1}{N_{j_A}}  \frac{1}{N_{j_B}} \text{ cov} \left(\sum_{i=1}^{N_{j_A}} X_{ik}, \sum_{i=1}^{N_{j_B}} X_{ik} \right)       \nonumber \\
    & = \frac{1}{\sigma^2_k} \sqrt{\frac{1}{N_{j_A}}} \sqrt{\frac{1}{N_{j_B}}} \sum_{i=1}^{N_{j_A}} \text{ cov} \left( X_{ik} , X_{ik}   \right)       \nonumber \\
    & = \frac{1}{\sigma^2_k} \sqrt{\frac{1}{N_{j_A}}} \sqrt{\frac{1}{N_{j_B}}} N_{j_A} \sigma^2_k \nonumber \\
    & = \sqrt{\frac{N_{j_A}}{N_{j_B}}}
\end{align}

For a single stage $j$, the correlation coefficient between two test statistics for outcomes $k_1, k_2, \, k_1 \neq k_2$ is $\rho_{k_1k_2}$. The covariance cov$(Z_{jk_1}, Z_{jk_2})$ is then

\begin{align}\label{eq:one_stage}
    \text{ cov}(Z_{jk_1}, Z_{jk_2}) & = \text{ cov} \left( \frac{\hat{\mu}_{k_1}}{\sqrt{\sigma^2_{k_1}/n_j}} , \frac{\hat{\mu}_{k_2}}{\sqrt{\sigma^2_{k_2}/n_j}}  \right) \nonumber \\
    & = \sqrt{\frac{n_j}{\sigma^2_{k_1}}}  \sqrt{\frac{n_j}{\sigma^2_{k_2}}} \text{ cov} \left(\hat{\mu}_{k_1}, \hat{\mu}_{k_2}\right)   \nonumber \\
    & = \frac{n_j}{\sqrt{\sigma^2_{k_1} \sigma^2_{k_2}}} \text{ cov} \left( \frac{1}{n_j} \sum_{i=1}^{n_j} X_{ik_1} ,  \frac{1}{n_j} \sum_{i=1}^{n_j} X_{ik_2}   \right) \nonumber \\
    & = \frac{1}{n_j \sqrt{\sigma^2_{k_1} \sigma^2_{k_2}}} \text{ cov} \left( \sum_{i=1}^{n_j} X_{ik_1} , \sum_{i=1}^{n_j} X_{ik_2}   \right) \nonumber \\
    & = \frac{1}{n_j \sqrt{\sigma^2_{k_1} \sigma^2_{k_2}}}  \sum_{i=1}^{n_j} \text{ cov} \left( X_{ik_1}, X_{ik_2} \right) \nonumber \\
    & = \frac{1}{n_j \sqrt{\sigma^2_{k_1} \sigma^2_{k_2}}} n_j \rho_{k_1 k_2}  \sigma_{k_1} \sigma_{k_2} \nonumber \\
    & = \rho_{k_1 k_2}   
\end{align}

The covariance of two test statistics for stages $j_A, j_B, \, j_B \geq j_A$ and outcomes $k_1, k_2, \, k_1 \neq k_2$, is

\begin{align}\label{eq:two_stages_two_outcomes}
    \text{cov}\left( Z_{j_A k_1}, Z_{j_B k_2} \right) & = \text{ cov}\left( \sqrt{\frac{N_{j_A}}{\sigma^2_{k_1}}} \hat{\mu}_{j_A k_1} ,  \sqrt{\frac{N_{j_B}}{\sigma^2_{k_2}}} \hat{\mu}_{j_B k_2} \right)  \nonumber \\
    & = \sqrt{\frac{N_{j_A}}{\sigma^2_{k_1}}}  \sqrt{\frac{N_{j_B}}{\sigma^2_{k_2}}}  \text{ cov}\left( \hat{\mu}_{j_A k_1},  \hat{\mu}_{j_B k_2}  \right)  \nonumber \\
    & = \sqrt{\frac{N_{j_A}}{\sigma^2_{k_1}}}  \sqrt{\frac{N_{j_B}}{\sigma^2_{k_2}}}  \text{ cov}\left( 
     \frac{1}{N_{j_A}} \sum_{i=1}^{N_{j_A}} X_{i k_1} , \frac{1}{N_{j_B}} \sum_{i=1}^{N_{j_B}} X_{i k_2}  \right)  \nonumber \\
    & =  \sqrt{\frac{N_{j_A}}{\sigma^2_{k_1}}}  \sqrt{\frac{N_{j_B}}{\sigma^2_{k_2}}} \frac{1}{N_{j_A}} \frac{1}{N_{j_B}}  \text{ cov}\left( \sum_{i=1}^{N_{j_A}} X_{i k_1} , \sum_{i=1}^{N_{j_B}} X_{i k_2}  \right)  \nonumber \\
    & = \frac{1}{\sqrt{\sigma^2_{k_1} \sigma^2_{k_2} N_{j_A} N_{j_B} }}  \sum_{i=1}^{N_{j_A}}  \text{ cov}\left( X_{i k_1}, X_{i k_2}  \right)  \nonumber \\
    & = \frac{1}{\sqrt{\sigma^2_{k_1} \sigma^2_{k_2} N_{j_A} N_{j_B} }} N_{j_A} \rho_{k_1 k_2} \sigma_{k_1}  \sigma_{k_2}   \nonumber \\
    & = \sqrt{\frac{N_{j_A}}{N_{j_B}}} \rho_{k_1 k_2}
\end{align}

Combining Equations~(\ref{eq:one_outcome}), (\ref{eq:one_stage}) and (\ref{eq:two_stages_two_outcomes}), the covariance cov$(Z_{j_A k_1}, Z_{j_B k_2})$ for any $j_A, j_B, \, j_B \geq j_A$ and $k_1, k_2$ can be stated as

\begin{equation}\label{eq:covariance}
  \text{cov}(Z_{j_A k_1}, Z_{j_B k_2}) = \left\{
  \begin{array}{@{}ll@{}}
    1                              & \text{if } j_A = j_B \text{ and }  k_1 = k_2  \\[10pt]
    \rho_{k_1 k_2}                 & \text{if } j_A = j_B \text{ and } k_1 \neq k_2 \\[10pt]
    \sqrt{\frac{N_{j_A}}{N_{j_B}}} & \text{if } j_A\neq j_B\text{ and } k_1 = k_2 \\[10pt]
    \rho_{k_1 k_2} \sqrt{\frac{N_{j_A}}{N_{j_B}}} & \text{if } j_A\neq j_B\text{ and } k_1 \neq k_2 \\[10pt]
  \end{array}\right\}.
\end{equation} 

This allows the construction of a covariance matrix for test statistics, for any number of stages $J$ and outcomes $K$. This covariance matrix is necessary to describe the multivariate normal distribution of the test statistics, shown in Equation~(\ref{eq:test_statistics_MVN}) directly below. Note: in this equation, each element of the covariance matrix cov$(Z_{jk}, Z_{jk})$ is presented simply as $jk, jk$ to save space.

\begin{landscape}
\begin{align}
\label{eq:test_statistics_MVN}
    \begin{pmatrix}
    Z_{j_1 k_1} \\
    Z_{j_1 k_2} \\
    \vdots \\
    Z_{j_1 K} \\
    Z_{j_2 k_1} \\
    Z_{j_2 k_2} \\
    \vdots \\
    \vdots \\
    Z_{JK}
    \end{pmatrix}
& \sim MVN_{JK}
\begin{pmatrix}
    \begin{pmatrix}
    \hat{\tau}_{j_1 k_1} \sqrt{\frac{N_{j_1}}{\sigma^2_{k_1}}}\\
    \hat{\tau}_{j_1 k_2} \sqrt{\frac{N_{j_1}}{\sigma^2_{k_2}}}\\
    \vdots\\
    \hat{\tau}_{j_1K} \sqrt{\frac{N_{j_1}}{\sigma^2_K}}\\
    \hat{\tau}_{j_2 k_1} \sqrt{\frac{N_{j_2}}{\sigma^2_{k_1}}}\\
    \hat{\tau}_{j_2 k_2} \sqrt{\frac{N_{j_2}}{\sigma^2_{k_2}}}\\
    \vdots\\
    \vdots\\
    \hat{\tau}_{JK} \sqrt{\frac{N_J}{\sigma^2_K}}\\
    \end{pmatrix}
    ,
    \begin{pmatrix}
        j_1k_1,j_1k_1 & j_1k_1,j_1k_2 & \hdots & j_1k_1,j_1K & j_1k_1,j_2k_1 & j_1k_1,j_2k_2 & \hdots & \hdots & j_1k_1,JK\\
        j_1k_2,j_1k_1 & j_1k_2,j_1k_2 & \hdots & j_1k_2,j_1K & j_1k_2,j_2k_1 & j_1k_2,j_2k_2 & \hdots & \hdots & j_1k_2,JK\\
        \vdots & \vdots & \ddots & \vdots & \vdots & \vdots & \ddots & \ddots & \vdots\\
        j_1K,j_1k_1   & j_1K,j_1k_2   & \hdots & j_1K,j_1K & j_1K,j_2k_1 & j_1K,j_2k_2 & \hdots & \hdots & j_1K,JK\\
        j_2k_1,j_1k_1 & j_2k_1,j_1k_2 & \hdots & j_2k_1,j_1K & j_2k_1,j_2k_1 & j_2k_1,j_2k_2 & \hdots & \hdots & j_2k_1,JK\\
        j_2k_2,j_1k_1 & j_2k_2,j_1k_2 & \hdots & j_2k_2,j_1K & j_2k_2,j_2k_1 & j_2k_2,j_2k_2 & \hdots & \hdots & j_2k_2,JK\\
    \vdots & \vdots & \ddots & \vdots & \vdots & \vdots & \ddots & \ddots & \vdots\\
    \vdots & \vdots & \ddots & \vdots & \vdots & \vdots & \ddots & \ddots & \vdots\\
        JK,j_1k_1 & JK,j_1k_2 & \hdots & JK,j_1k_1 & JK,j_1K & JK,j_2k_1 & \hdots & \hdots & JK,JK\\
    \end{pmatrix}
\end{pmatrix}
\end{align}
\end{landscape}

\subsection{Type-I error-rate and power}\label{sec:MO_typeIerror_power}

Define $R(\boldsymbol{\mu} \vert K, m, J, C, \Delta)$ as the probability of rejecting the null hypothesis when the true outcome effects are equal to $\boldsymbol{\mu}=(\mu_1, \mu_2, \dots, \mu_K)$, for some design realisation characterised by $K, m, J, C$ and $\Delta$. The probability $R(\boldsymbol{\mu} \vert K, m, J, C, \Delta)$ can be readily evaluated using simulation (described in Section~\ref{sec:MO_simulation_integration}). We define type-I error-rate as 
\[
\alpha^* = R(\boldsymbol{\mu} = \mathbf{0} \vert K, m, J, C, \Delta).
\]

That is, we control the type-I error-rate under the scenario where $\boldsymbol{\mu} = \mathbf{0}$. Cook and Farewell~\cite{cook1994guidelines} have previously used this manner of type-I error control in a multiple outcome setting. This is in contrast to Lehmann and Romano~\cite{lehmann2012generalizations}, who treat each hypothesis separately and describe controlling the probability of rejecting $k$ true hypotheses as the $k$-familywise error rate (where $k \equiv m$ here). Dmitrienko et al.~\cite{dmitrienko2009multiple} refer to this as the generalized familywise error rate while Grayling et al.~\cite{grayling2017efficient} describe this as the $a$-generalised type-I familywise error rate. Similarly, we define power as 
\[
1-\beta^* = R\left( \boldsymbol{\mu} = \boldsymbol{\delta}_\beta \vert K, m, J, C, \Delta \right),
\]

for some vector of effect sizes $\boldsymbol{\delta}_\beta = (\delta_{\beta 1}, \delta_{\beta 2}, \dots, \delta_{\beta K})$ for which we would like to control the probability of rejecting $H_0$. 

Let the required type I and type II errors be $\alpha$ and $\beta$. We require designs that satisfy the conditions $\alpha^* = R\left(\boldsymbol{\mu} = \mathbf{0} \vert K, m, J, C, \Delta \right) \leq \alpha$ and $1-\beta^* = R\left( \boldsymbol{\mu} = \boldsymbol{\delta}_\beta \vert K, m, J, C, \Delta\right) \geq 1-\beta$. The stopping boundaries $f_j, e_j$ are then determined by one-dimensional optimisation to find the value of $C$ that minimises $(\alpha-\alpha^*)^2$. With $\alpha^*$ obtained and $C$ fixed, $1-\beta^*$ is found for some small initial $n$, which is increased until the required power is reached.

Though we choose to control type-I error-rate and power at one particular point each,  $R(\boldsymbol{\mu} \vert K, m, J, C, \Delta) \leq \alpha$ and $R(\boldsymbol{\mu} \vert K, m, J, C, \Delta) \geq 1-\beta$ for (two different) $K$-dimensional regions. One may be interested in not only controlling type-I error-rate and power at a single point, but across certain regions. This idea is explored further in Section~\ref{sec:MO_rejection_regions}.

With regards to powering the trial for a certain point $\boldsymbol{\delta}_\beta$, we specify anticipated lower and greater effect sizes for each outcome, $\boldsymbol{\delta_0}=(\delta_{01}, \delta_{02}, \dots, \delta_{0K})$ and $\boldsymbol{\delta_1}=(\delta_{11}, \delta_{12}, \dots, \delta_{1K})$. We then set $\boldsymbol{\delta}_\beta = (\delta_{11}, \dots, \delta_{1m}, \delta_{0(m+1)}, \dots, \delta_{0K}).$ That is, exactly $m$ outcomes are equal to their greater anticipated effect $\delta_{1k}$, while $K-m$ outcomes are equal to their lower anticipated effect $\delta_{0k}$. This is analogous to the least favourable configuration (LFC) described by Thall et al.~\cite{Thall1989_LFC} in the context of multi-arm trials. In such trials, the probability of correctly concluding not only that a promising treatment exists, but also identifying that treatment, is of obvious importance. However, in the context of a single-arm trial with multiple outcomes, we place prime importance on the probability of correctly concluding that some subset of $m$ or more outcomes show promise, rather than additionally correctly identifying the outcomes in this subset. In some situations, it may be of great importance to correctly identify the outcomes that show promise. If so, we can redefine power as the probability of both rejecting the null hypothesis when at least $m$ outcomes have a promising effect size and correctly identifying $m$ of those outcomes. 

Above, the $m$ ``working'' outcomes are taken to be simply the first $m$ outcomes, without loss of generality. They may alternatively be set to be the $m$ smallest standardised outcome effects $\delta_{1k}/\sigma_k, k=\, \dots, K$. This may be of use when the anticipated outcome effects, or anticipated variances, differ. In such a case, it would be desirable to power a trial to correctly conclude that $m$ outcomes show promise when such promising outcomes have the $m$ smallest standardised anticipated effects; in single-outcome trials, identifying small effects requires a larger sample size than identifying large effects, and so power is minimised when the $m$ promising outcomes are those with the $m$ smallest standardised effect sizes.

\subsection{Integration vs. simulation}\label{sec:MO_simulation_integration}
For both multi-outcome approaches, simulation rather than integration is used to obtain design realisations and their operating characteristics. Grayling et al.~\cite{grayling2017efficient} present the following notation that fully characterises the progress and conclusion of a MAMS design based on $K$ outcome-specific hypotheses $H_k, \, 1, \dots K$: $\boldsymbol{\Psi}=(\Psi_1, \Psi_2, \dots, \Psi_K)$, $\boldsymbol{\Omega}=(\omega_1, \omega_2, \dots, \omega_K)$, where $\Psi_k=1$ if $H_k$ is rejected, $\Psi_k=0$ otherwise and $\Omega_k=j$ where $j$ is either the stage at which $H_k$ is rejected or not rejected or where the trial is stopped. In our multi-outcome multi-stage approach, the test statistics of all outcomes at stage $j$ must considered simultaneously. There are no outcome-specific hypotheses that may be rejected independently of others. It is not sufficient to know that an outcome has crossed a boundary: an outcome may cross a boundary and no trial decision is taken. It is necessary to know the \textit{state} of each outcome's test statistic, that is, which boundary it has crossed (if any), at every stage. As such, it is not possible in this approach to characterise a trial's progress using two $K$-length vectors. What is required is $J$ $K$-length vectors or a $J \times K$ matrix, for example:

\[
P = 
\begin{bmatrix}
\Psi_{11} & \Psi_{12} & \dots  & \Psi_{1K}\\
\Psi_{21} & \ddots    &        & \vdots   \\
\vdots    &           & \ddots & \vdots   \\
\Psi_{J1} & \hdots    & \hdots & \Psi_{JK}
\end{bmatrix}
, \;
\text{ where }
\Psi_{jk} = 
  \begin{cases}
    1 & \text{if } Z_{jk} > e_j \\
    0 & \text{if } e_j \geq Z_{jk} \geq f_j \\
    -1 & \text{if } Z_{jk} < f_j
  \end{cases}
.
\]

In this matrix, each row represents a single stage. As in Grayling et al.~\cite{grayling2017efficient}, the probability of a particular instance of trial progress can be found through a $JK$-dimensional integration. Each $\Psi_{jk} = \{ -1, 0, 1 \}$ as defined above has three possible states, and so there are a maximum of $3^{JK}$ possibilities for the progress of the trial, akin to the ``paths'' of binary outcome trials described by Law et al.~\cite{law}, and the probability of each can be calculated using the corresponding $JK$-dimensional integration. The number of possibilities of interest, and so the number of $JK$ integrations required, can be reduced from $3^{JK}$. For example, the probability that a trial will end at the first stage need only consider possible states at stage 1. Other reductions are possible, but the degree of reductions required may need to be considerable to manage even a modest trial of of $J=3$ stages and $K=3$ outcomes ($3^9 = 19683$ multiple integrals). Conversely, on a computer with an i7-3770 processor and 16GB RAM with no parallelisation, it is possible to simulate $10^5$ multi-outcome multi-stage trials of our approach in under 10 seconds.

\subsection{Design search}\label{sec:mo_design_search}
We seek to obtain the design realisation that minimises $N$ while satisfying the required type-I error-rate and power. As stated above, the design search for this approach uses simulation. Specifically, we simulate aggregated trial results by simulating $JK$ test statistics, representing the test statistic at each stage $j$ and for each outcome $k$. We simulate from a multivariate normal distribution consisting of a mean that is a null vector of length $JK$ and the covariance matrix in Equation~(\ref{eq:test_statistics_MVN}). This approach was suggested by Wason and Jaki~\cite{wason2012optimal}.

The remaining components of the design search for this approach are described using pseudocode below. Briefly, an optimiser is used in conjunction with Algorithm~\ref{algo:MO_pReject} to find the constant $C$ and corresponding set of lower and upper boundaries that minimise $(\alpha-\alpha^*)^2$. This means that the final design will have a type-I error-rate $\alpha^* \approx \alpha$. To strictly ensure $\alpha^* \leq \alpha$, one might choose to find boundaries by minimising the discontinuous function $(\alpha-\alpha^*)^2$ if $\alpha^* \leq \alpha$, 1 if $\alpha^* > \alpha$. 
With $\alpha^*$ obtained and $C$ fixed, $1-\beta^*$ is found for some small initial $n$, which is increased until the required power is reached. This is shown in Algorithm~\ref{algo:MO_find_N}.

\begin{algorithm}[htbp!]
\SetAlgoLined
lower.bounds $\leftarrow$ call findLowerBounds($C, J, \Delta$)\;
upper.bounds $\leftarrow$ call findUpperBounds($C, J, \Delta$)\;
nsims $\leftarrow$ nrow($TS$) \;
\For{each $i$ in 1 to nsims}{
    \For{each $j$ in 1 to $J$}{
    TS.current.stage $\leftarrow$ call findCurrentTSStage($TS, i, J,K$)\;
    \eIf{sum(TS.current.stage $>$ upper.bounds[$j$]) $\geq m$ }{
    go[$i, j$] $\leftarrow$ 1\;
    nogo[$i, j$] $\leftarrow$ 0\;
    }{\eIf{sum(TS.current.stage $<$ lower.bounds[$j$]) $\geq K-m+1$}{
    nogo[$i, j$] $\leftarrow$ 1\;
    go[$i, j$] $\leftarrow$ 0\;
    }
    {
    go[$i,j$] $\leftarrow$ 0\;
    nogo[$i, j$] $\leftarrow$ 0\;
    }
    }
    }
    go.nogo.decision[$i$] $\leftarrow$ call findEarliestDecision(go[$i$, ], nogo[$i$, ]) \;
    stop.stage[$i$] $\leftarrow$ call findStageOfEarliestDecision(go[$i$, ], nogo[$i$, ])\;
}
go.decision.count $\leftarrow$ sum(go.nogo.decision)==``go'' \;
p.reject.null $\leftarrow$ go.decision.count/nsims \;
expected.stages.count $\leftarrow$ sum(stop.stage)/nsims\;
\If{exists($\alpha$)}{
value.to.minimise $\leftarrow$ (p.reject.null $- \alpha$)$^2$\;
}
\caption{Function pReject: for finding $R()$ and expected number of stages. Input: $C, J, K$, $m, \Delta, \alpha, TS \text{ (matrix of test statistics)}$}
\label{algo:MO_pReject}
\end{algorithm}

\begin{algorithm}[htbp!]
\SetAlgoLined
$C \leftarrow$ call optimise(pReject($J, K, m, \Delta, \alpha$, $TS$))\;
typeIerror $\leftarrow$ call pReject($C, J, K,m, \Delta, \alpha, TS$)\;
$\boldsymbol{\mu} \leftarrow \boldsymbol{\delta}_\beta$ \tcp*[f]{mu can be set to any vector, if another definition of power is desired}\;
pow $\leftarrow$ 0\;
n.current $\leftarrow$ nmin-1\;
\While{pow $< 1-\beta$}{
    n.current $\leftarrow$ n.current+1\;
    $\mathcal{I} \leftarrow$ call findInformation(current.n, $\boldsymbol{\sigma}$)\;
    $\boldsymbol{\tau} \leftarrow$ call findEffects($\boldsymbol{\mu}, \mathcal{I}$)\;
    TS.current $\leftarrow$ call addEffectsToTS($\boldsymbol{\tau}, TS$)\;
    pow $\leftarrow$ call pReject($C$, n.current, TS.current, $1-\beta$)
}
\caption{Find optimal $C$ for type-I error-rate control, then find smallest value of $N$ that satisfies $1-\beta^* \geq 1 - \beta$. Input: $J, K, m, \Delta, \alpha, TS, \boldsymbol{\delta}_\beta, \beta, \boldsymbol{\sigma}$, nmin.}
\label{algo:MO_find_N}
\end{algorithm}

\subsection{Composite outcome design}
A simple composite outcome can be created at each stage $j$ by summing the $K$ test statistics $Z_{jk}$. Let the composite test statistic at stage $j$ be $Z_j=\sum\limits_{k=1}^K Z_{jk}$. Each $Z_{jk}$ has been standardised (see Section~\ref{sec:MOMS_design_1}), therefore each $Z_j$ is standardised. By taking the sum of the outcomes, all outcomes are being weighted equally. An investigator may choose to apply unequal weights to the outcomes. We undertake a design search analogous to the multi-outcome design search described above, again to find the design realisation that satisfies type-I error-rate and power while minimising $N$. The same simulated data is used, with the test statistics for each outcome summed to create a composite test statistic for each stage $j$ as described. Again an optimiser is used in conjunction with Algorithm~\ref{algo:MO_pReject} to find some constant $C_{COMP}$ and corresponding stopping boundaries that result in an acceptable type-I error-rate, that is, $\alpha^* \leq \alpha$. The procedure in Algorithm~\ref{algo:MO_find_N} is then used to find the smallest sample size $N$ that will result in an acceptable power, that is, $1-\beta^* \geq 1-\beta$.

\subsection{Comparing multi-outcome and composite designs}\label{sec:mo_comparing_des_part1}
The multi-outcome and composite approaches were compared by obtaining design realisations that satisfied the required type-I error-rate and power, set at $\alpha=0.025$ and $1-\beta = 0.8$. These $\alpha$ and $1-\beta$ were chosen to align with those used in Sozu et al.~\cite{Sozu}. The anticipated outcome effect sizes were set as $\delta_{01}=\delta_{02}= \dots = \delta_{0K}=\delta_0=0.2$ and $\delta_{11}=\delta_{12}= \dots = \delta_{1K}=\delta_1=0.4$, again in alignment with Sozu et al., and $\boldsymbol{\delta}_\beta = (\delta_{11}, \dots, \delta_{1m}, \delta_{0(m+1)}, \dots, \delta_{0K}).$ as described in Section~\ref{sec:MO_typeIerror_power}. For simplicity, the variance of each outcome is fixed and equal to one, that is, $\sigma^2_k=\sigma^2=1, \, \forall k$. $\Delta=0$ is used in the calculation of stopping boundaries, equivalent to the stopping boundaries proposed by O'Brien and Fleming~\cite{obrien1979multiple}. The reported operating characteristics are the probability of rejecting the null hypothesis and ESS under the LFC.

We firstly compare rejection regions for single-stage multi-outcome and composite designs. This is followed by comparing design realisations for varying values of correlation $\rho$. Correlation $\rho_{k_1k_2}=\rho, \, k_1 \neq k_2$ between all outcomes was equal, and the values examined were $\rho \in \{0, 0.1, \dots, 0.8\}$.

It is also of interest to examine the consequences of specifying different true outcome effects, given some anticipated outcome effects $\boldsymbol{\delta_\beta}$. When the true effect sizes differ from the effect sizes anticipated in the designs, the performance of both designs will be affected. While we may anticipate which approach may perform better under certain conditions, we wish to quantify these relative changes in performance. We therefore search for design realisations as described above, for both multi-outcome and composite approaches, and note the effect of changing the true outcome effects $\boldsymbol{\mu}$. The required type-I error-rate and power and anticipated outcome effect sizes specified above $(\alpha, \beta, \delta_0, \delta_1, \boldsymbol{\delta}_\beta)$ were also used here, with a shared correlation $\rho_{k_1k_2}=\rho=0.3, \, k_1 \neq k_2$.

\section{Results: Multi-outcome multi-stage design with general number of required efficacious outcomes}

\subsection{Comparison of single-stage rejection regions}\label{sec:MO_rejection_regions}
\begin{figure}[hbtp!]
\centering
\includegraphics[width=\textwidth]{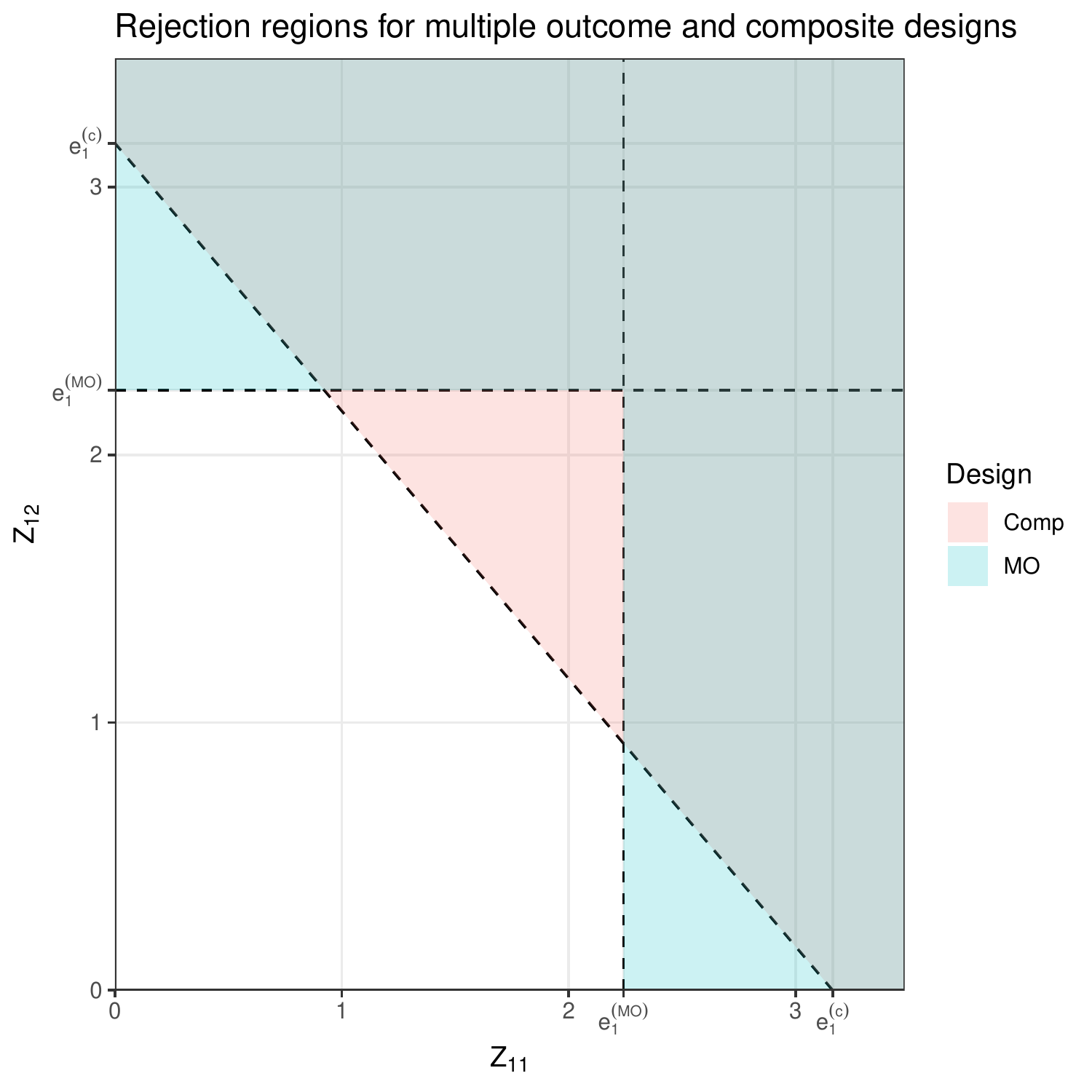}
\caption{Comparison of final rejection regions for multi-outcome design (blue) and composite design (red), for $\{K=2, m=1, J=1\}$.}
\label{fig:rejection_region}
\end{figure}

The multi-outcome and composite design approaches lead to different rejection regions. An example of this is shown in Figure~\ref{fig:rejection_region}, where a design realisation for each approach has been obtained and the final rejection regions overlaid. The outcome design parameters were $\{K=2, m=1, J=1\}$, that is, single-stage designs. For a composite design, let the lower and upper stopping boundaries for a trial of $j$ stages be $\mathbf{f}^{(c)} = (f_1^{(c)}, f_2^{(c)}, \dots, f_J^{(c)})$ and $\mathbf{e}^{(c)} = (e_1^{(c)}, e_2^{(c)}, \dots, e_J^{(c)})$ respectively. For this particular composite design, where $K=2, J=1$, the null hypothesis will be rejected at the end of the trial \textit{iff} the sum of the test statistics $Z_{11}, Z_{12}$ is greater than some corresponding efficacy boundary $e_1^{(c)}$, or in general, $\left( \sum_{k=1}^K Z_{Jk} \right) > e_J^{(c)}$. For this particular multi-outcome design, the null hypothesis will be rejected at the end of the trial \textit{iff} either test statistic exceeds some corresponding efficacy boundary $e_1$, or in general, $\left( \sum_{k=1}^K \mathbb{I}(Z_{Jk} > e_J) \right) \geq m$. Thus for a general number of outcomes $K$, rejection of the null hypothesis using the composite design is dependent on all $K$ outcome test statistics, while rejection of the null hypothesis using the multiple-outcome design occurs if the test statistics of any $m$ outcomes show sufficient response.

The true effect sizes of the outcomes may differ from those specified in the design, and the nature of these differences may affect the performance of the designs in different ways. For example, we expect the multi-outcome approach to outperform the composite approach when some outcomes have a harmful ($\mu_k < 0$) effect, as these outcome effects will dilute any positive effects observed on the remaining outcomes. The opposite effect may occur when more than $m$ outcomes have some moderate effect. In this case, these moderate effects may combine under the composite design to increase the probability of rejecting null hypothesis compared to the multi-outcome design. We also expect the multi-outcome approach to perform better than the composite approach when fewer than $m$ outcomes have large effect sizes, as the additive aspect of the composite design may cause these outcomes' effects to outweigh the lack of effects in the remaining outcomes, again increasing the probability of rejecting null hypothesis compared to the multi-outcome design. Conversely, we expect the composite approach to perform better when more than $m$ outcomes true effect sizes at least as great as those anticipated, for the same reason. In this case, the outcomes' large effects would make correct rejection of the null hypothesis more likely.

\subsection{Varying correlation}
\begin{figure}[htbp!]
\centering
\includegraphics[width=0.8\textwidth]{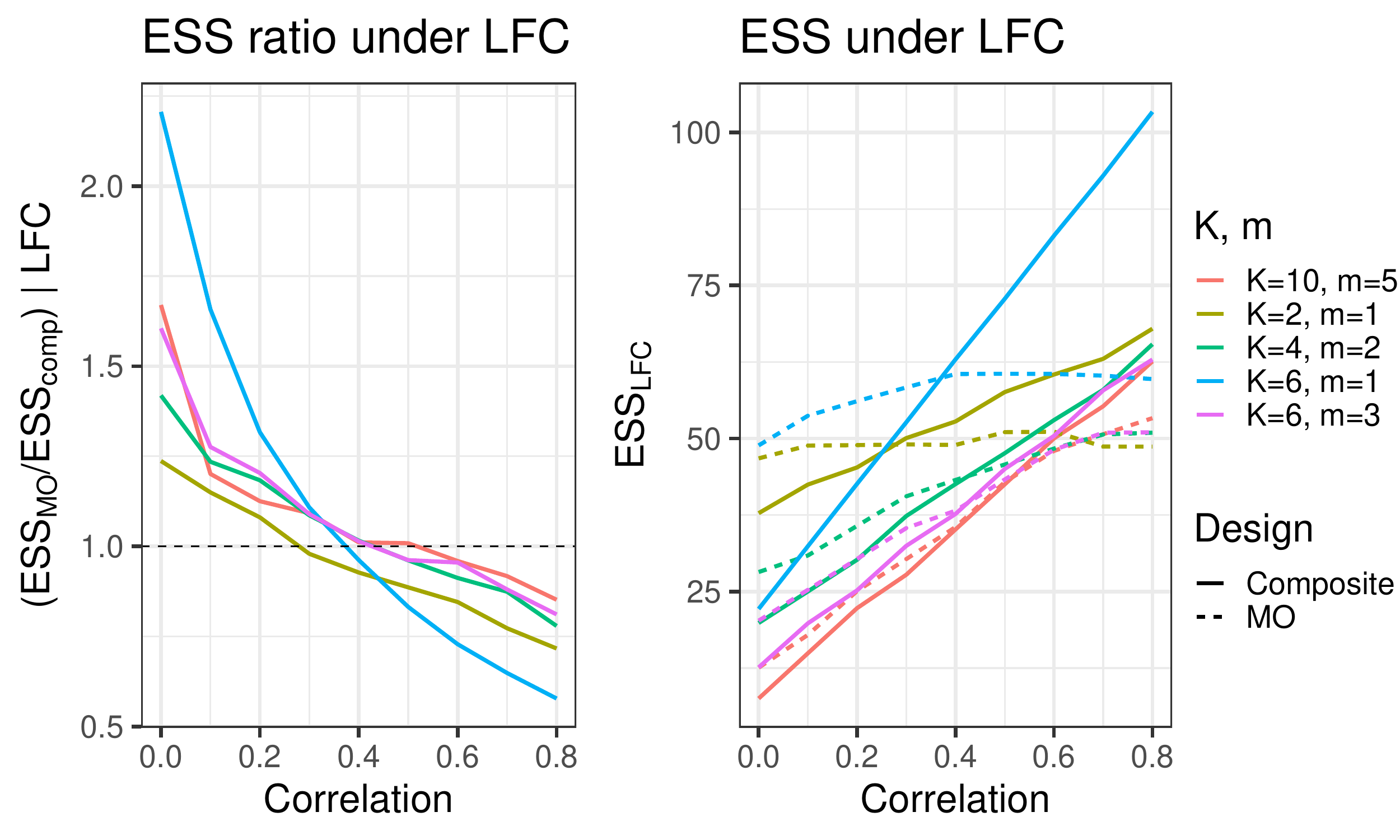}
\caption{Change in $ESS_{MO}/ESS_{comp}$ as correlation varies. $\alpha=0.025$, $1-\beta=0.8$, design parameters $J=3, \boldsymbol{\delta}_0=0.2, \boldsymbol{\delta}_1=0.4$. Simulations: $10^5$.}
\label{fig:MO_cor}
\end{figure}

\begin{figure}[htbp!]
    \centering
    \begin{subfigure}[b]{0.45\textwidth}
    \includegraphics[width=\textwidth]{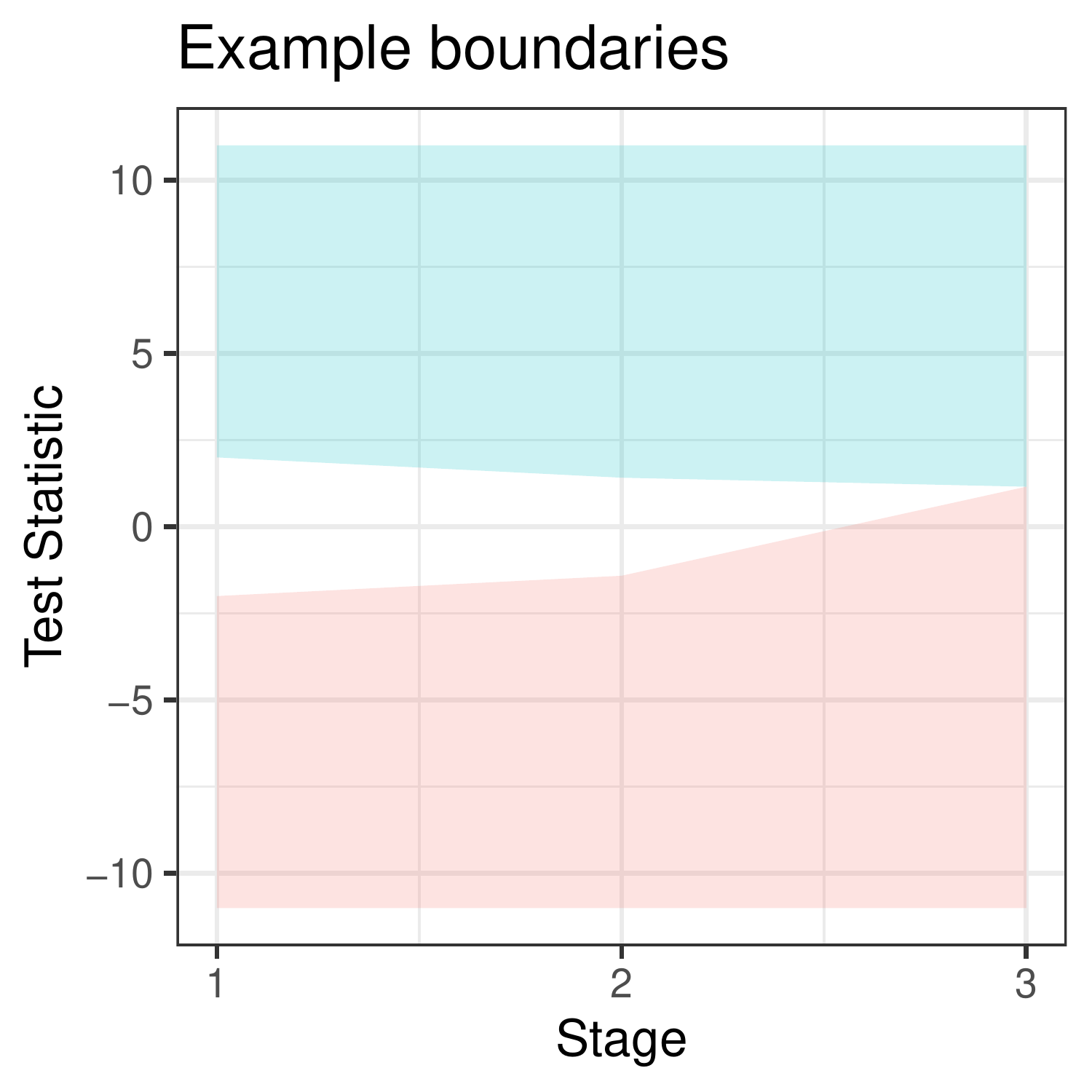}
    \caption{C=2}
    \label{fig:small_bounds}
    \end{subfigure}
    ~ 
    \begin{subfigure}[b]{0.45\textwidth}
    \includegraphics[width=\textwidth]{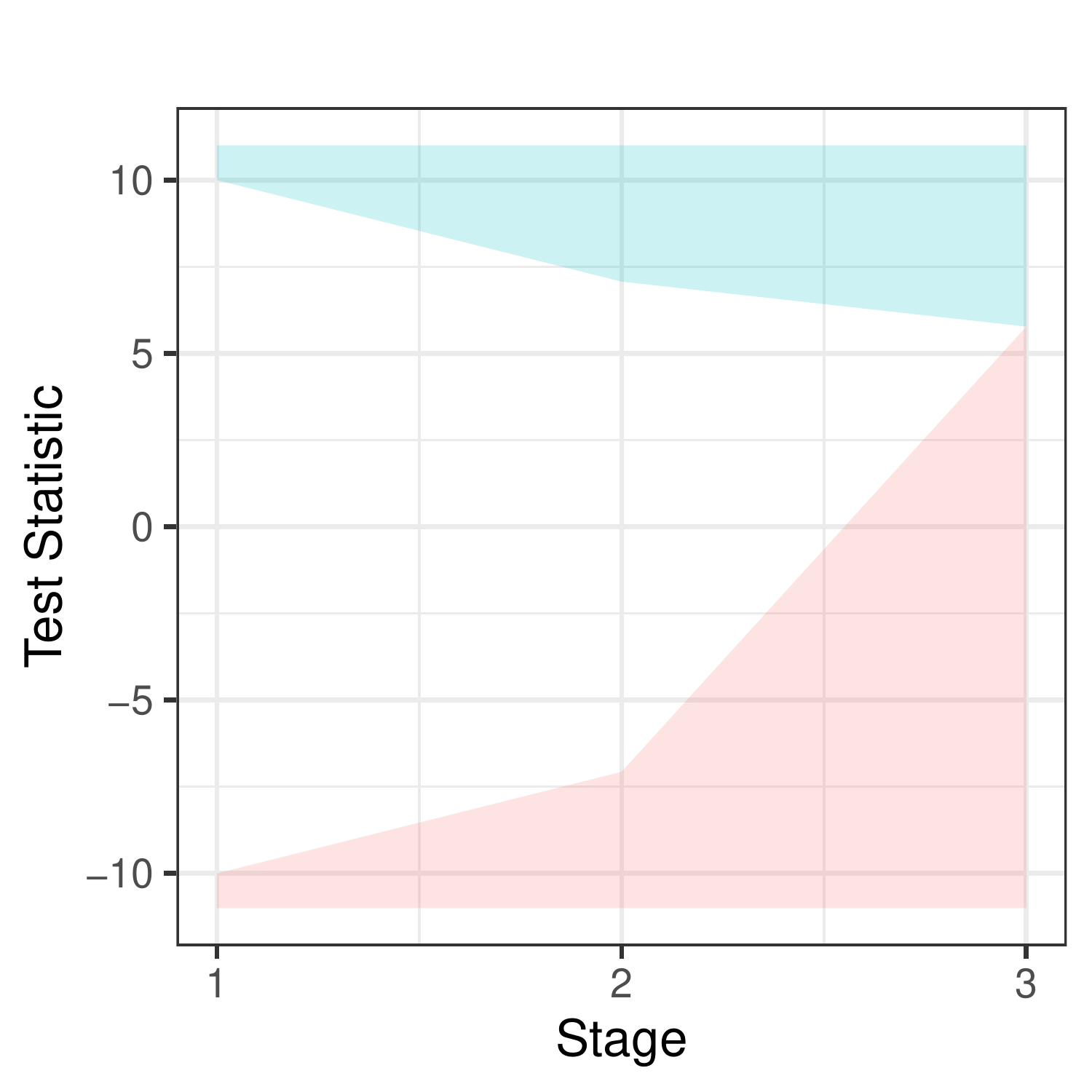}
    \caption{C=10}
    \label{fig:big_bounds}
    \end{subfigure}
    \caption{Examples of Wang and Tsiatis boundaries for three-stage trials, using $\Delta$=0 and $C=\{2, 10\}$.}
    \label{fig:compare_bounds}
\end{figure}

Figure~\ref{fig:MO_cor} compares the multi-outcome design to the composite design in terms of ESS under the LFC. Define ESS$_{MO}$ and ESS$_{comp}$ as the ESS under the LFC for the multi-outcome and composite designs respectively. The ESS ratio $ESS_{MO}/ESS_{comp}$ under the LFC is shown as correlation $\rho$ varies $(\rho \in \{0, 0.1, \dots, 0.8\})$. The number of stages was $J=3$, with the following sets of $\{K, m \}$: $\{K=2, m=1\}, \{K=4, m=2\}, \{K=6, m=1\}, \{K=6, m=3\}, \{K=10, m=5\}$ . A value of less than 1 means that the ESS under LFC is smaller for the multi-outcome design compared to the composite design. Also of interest is the ENM for a given design. In these two approaches, all $K$ outcomes are measured for $n$ participants at each stage $j$ that takes place. As such, ENM in both approaches is simply $K \times ESS$, and so $ESS_{MO}/ESS_{comp} = ENM_{MO}/ENM_{comp}$. ESS ratio decreases as correlation increases. This means that when correlation is low, ESS is relatively poorer on the multi-outcome design, while when correlation is high, ESS is relatively better on the multi-outcome design. The change in ESS ratio as correlation varies is overwhelmingly due to the change in ESS$_{comp}$ as correlation increases. While ESS increases with correlation for both approaches (Figure~\ref{fig:MO_cor}, right), the increase is greater for the composite design. For the composite designs found, as correlation increases, so too does the constant $C$ that determines the stopping boundaries. The boundaries are chosen to ensure the correct type-I error-rate. As the composite test statistic is the sum of the outcome test statistics, increased correlation between outcomes makes type-I errors more likely. Using more extreme upper boundaries counteracts this to ensure an appropriate type-I error-rate. However, using the boundaries of Wang and Tsiatis~\cite{Wang1987} means that extreme upper boundaries are accompanied by extreme lower boundaries. Two contrasting examples are shown in Figure~\ref{fig:compare_bounds}, using $\Delta=0$ as for all design searches in this manuscript. The maximum sample size $N$ chosen is the smallest $N$ that results in adequate power. However, with high upper boundaries resulting from having highly correlated outcome test statistics, $N$ must be increased to ensure the design has adequate power.

The disparity in ESS between the methods is greatest when $K=6, m=1$, the only instance where $m/K < 0.5$. The improvement in ESS under the multi-outcome design compared to the composite design as correlation increases is similar for the remaining combinations, where $m/K=0.5$. Among these combinations, those with a smaller number of outcomes $K$ appear to benefit more from using a multi-outcome approach compared to a composite approach. For these composite designs, stopping boundaries are independent of $m$, as type-I error-rate (which is driven by the stopping boundaries) is calculated under the global null. Therefore the boundaries for, say, $\{K=6, m=1\}$ and $\{K=6, m=3\}$ differ only due to simulation error. However, power is calculated under the LFC, $\boldsymbol{\mu} = \boldsymbol{\delta}_\beta$. As such, the observed outcome effects are greater as $m$ increases. When $m/K$ is small, for example, when $\{K=6, m=1\}$, rejecting $H_0$ is less likely, and so $N$ increases to compensate for the lack of power. This explains the larger sample size for the composite design using design parameters $\{K=6, m=1\}$. Furthermore, when correlation is high, the $K-m$ null effects are less likely to contribute enough to the composite test statistic to increase power, exacerbating the need for a larger sample size.

\subsection{Varying true outcome effects}
\begin{figure}[htbp!]
\centering
\includegraphics[width=0.8\textwidth]{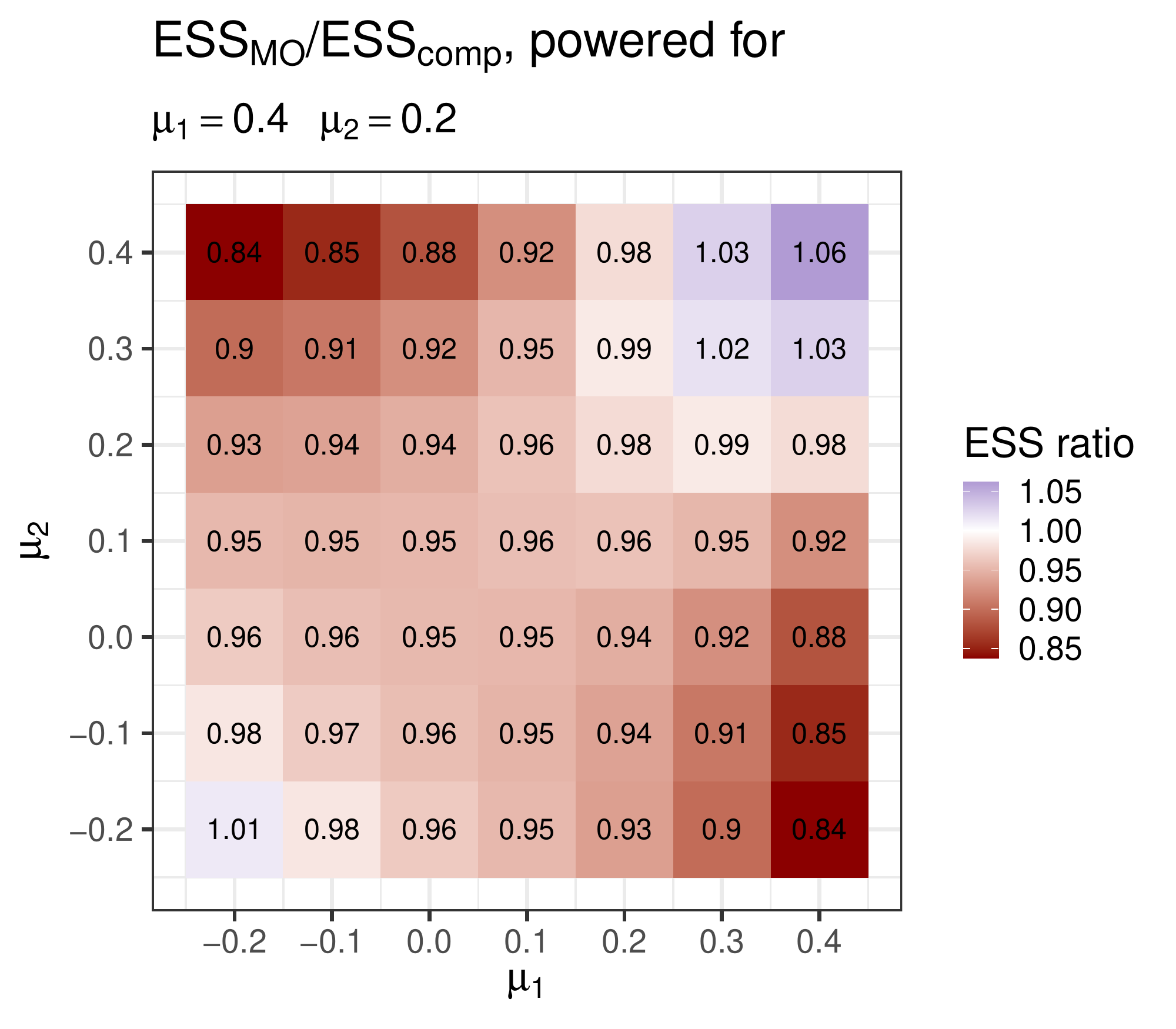}
\caption{Change in $ESS_{MO}/ESS_{comp}$ as true outcome effect sizes vary. Required type-I error-rate $\alpha=0.025$, power $1-\beta=0.8$, design parameters $\{K=2, m=1, J=3\}, \delta_{\beta 1}=0.4, \delta_{\beta 2}=0.2, \rho=0.3$. Simulations: $10^5$.}
\label{fig:MO_grid}
\end{figure}

In Figure~\ref{fig:MO_grid}, the ESS ratio is compared for a range of different true effects, for a single design realisation of each approach with $\{K=2, m=1, J=3\}$. In this case, the ESS ratio was obtained for every combination of true effect sizes $\mu_1, \mu_2 \in \{-0.2, -0.1, \dots, 0.4\}$, with anticipated effect sizes $\boldsymbol{\delta}_\beta=(0.4, 0.2)$ and design parameters $\alpha=0.025$, $1-\beta=0.8,$ $\rho_{k_1k_2}=\rho=0.3, \, k_1  \neq k_2$, $\sigma^2_k=\sigma^2=1, \, \forall k$. Using $K=2$ allows a grid of results to be plotted. Correlation $\rho=0.3$ is the point in Figure~\ref{fig:MO_cor} at which ESS ratio is close to one for $\{K=2, m=1\}$. The design realisations are $\{N=57, C=2.256490\}$ for the multi-outcome design and $\{N=60, C=3.240066\}$ for the composite design. Across all $(\mu_1, \mu_2)$ combinations in Figure~\ref{fig:MO_grid}, ESS is generally relatively lower when using the multi-outcome design, including the case where the true effect sizes are as anticipated $(\mu_1=0.4, \mu_2=0.2)$, though at this point the ESS ratio is close to one. The only regions where ESS is greater using the multi-outcome design is when the ``non-working'' outcome has a greater than anticipated effect size ($\mu_2 > 0.2$) and when both outcomes are particularly harmful ($\mu_1=-0.2, \mu_2=-0.2$). In the former case, the composite design is more likely to reject $H_0$ sooner as the design combines the positive observed effects of both outcomes. Similarly, in the latter case, the two negative observed effects combine, resulting in a test statistic that causes a trial to end for a no go decision sooner than the corresponding multi-outcome design.

\begin{figure}[htbp!]
\centering
\includegraphics[width=0.7\textwidth]{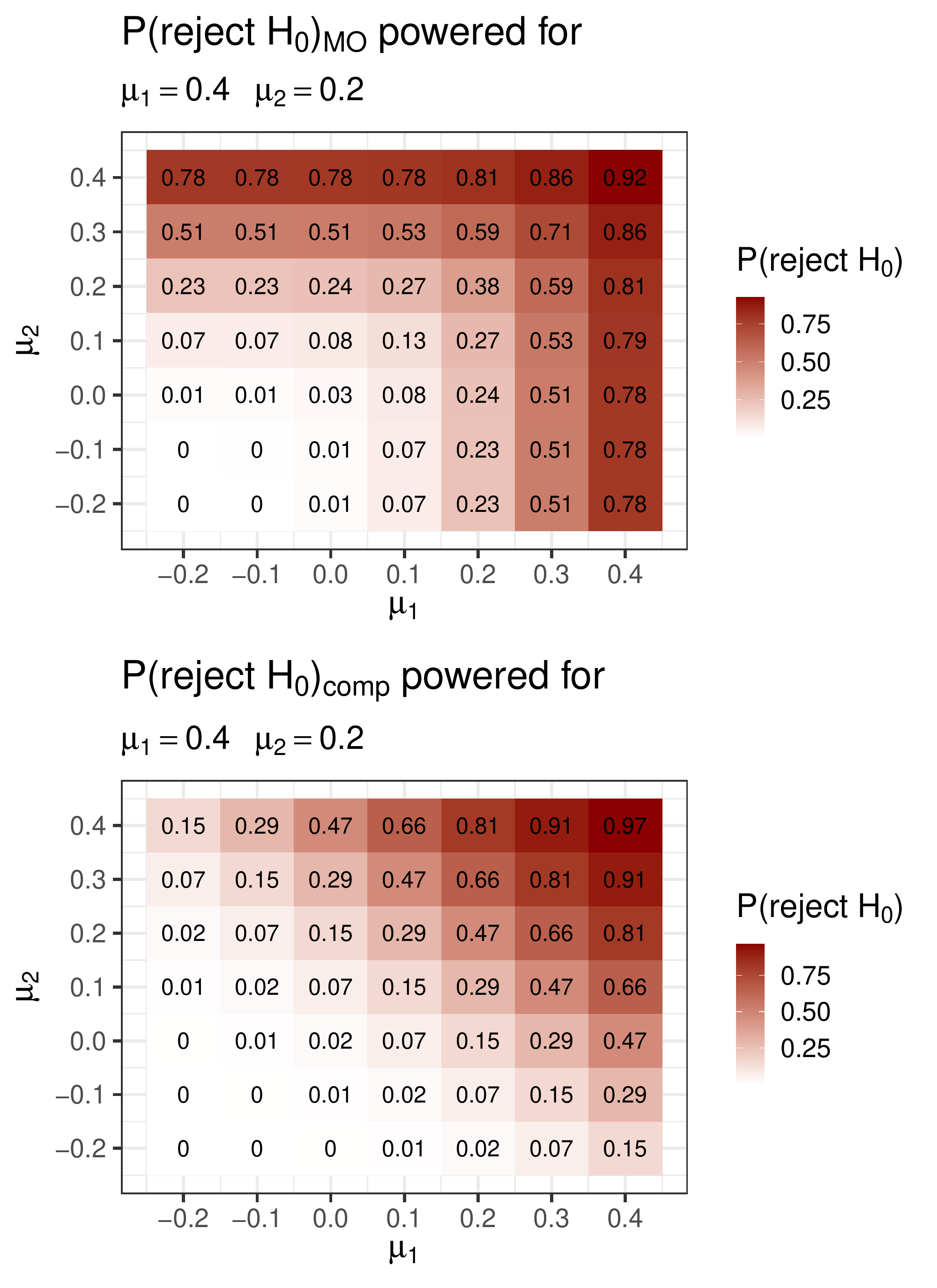}
\caption{$R(\boldsymbol{\mu}=\mu_1, \mu_2)$ as true outcome effect sizes vary. Required type-I error-rate $\alpha=0.025$, power $1-\beta=0.8$, design parameters $\{K=2, m=1, J=3\}$, $\boldsymbol{\delta}_\beta = (0.4, 0.2)$. Simulations: $10^5$.}
\label{fig:MO_preject_grid}
\end{figure}

Figure~\ref{fig:MO_preject_grid} shows the how the probability of rejecting $H_0$ changes for different true effects, for the same multi-outcome and composite design realisations as Figure~\ref{fig:MO_grid}. When using the multi-outcome design, P(reject $H_0$) remains at least close to the required power when either outcome has true effect $\mu=0.4$, while when using the composite design, P(reject $H_0$) decreases below the required power when one outcome has true effect $\mu=0.4$ and the other has some true effect less than 0.2. As above, the combining of outcome effects on the composite design is responsible for this, with the lower-than-anticipated observed effect ``cancelling out'' the positive observed effect to some extent. This can be seen in Figure~\ref{fig:rejection_region}, where a low value for test statistic $Z_{11} \text{ (or } Z_{12})$ means that a greater test statistic $Z_{12} \text{ (or } Z_{11})$ is required to reject $H_0$ under the composite design but not the multi-outcome design.

Table~\ref{tab:true_mo} also compares the two approaches in terms of a single design realisation for each approach, for a range of different true effects. In this case, the total number of outcomes is increased to $K=3$ while the remaining design parameters are unchanged. The design realisations are $\{N=60, C=2.394350\}$ for the multi-outcome design and $\{N=63, C=4.387731\}$ for the composite design. The ESS ratio is examined again, as is the probability of rejecting the null hypothesis, for a range of scenarios. The ESS ratio is greater than or equal to 1, i.e. ESS is poorer under the multi-outcome design, when all outcomes have equal non-zero true effects. Here, the composite design benefits from combining the observed effects. The ESS ratio is less than 1 otherwise, favouring the multi-outcome design. The relative difference in favour of the multi-outcome design is at its greatest when the ``non-working'' outcomes have a zero or harmful true effect, where the composite design either does not benefit or is even harmed by combining outcome effects. As in Figure~\ref{fig:MO_preject_grid}, under the multi-outcome design P(reject $H_0$) is close to the nominal power (or greater) when at least one outcome has a true effect equal to the anticipated effect, while under the composite design P(reject $H_0$) decreases as the true effect sizes of the ``non-working'' outcomes decrease, even if one outcome has a true effect equal to the anticipated effect. When all three outcomes have some true effect that is lower than $\delta_{1k}$, e.g. $\mu_1=\mu_2=\mu_3=0.3$ or $\mu_1=\mu_2=\mu_3=0.2$, rejecting the null hypothesis is more likely under the composite design than the multi-outcome design. Again, the multi-outcome design will only reject the null upon observing effects of a particular size on $m$ outcomes only, while the composite design may reach the rejection region by combining these smaller observed effects.

\begin{table}[htbp!]
\centering
\begin{tabular}{rrrrrrr}
  \toprule
$\mu_1$ & $\mu_2$ & $\mu_3$ & $R(\boldsymbol{\mu})_{MO}$ & $R(\boldsymbol{\mu})_{comp}$ & $ESS_{MO/comp}$ & Description \\ 
  \midrule
0.4 & 0.4 & 0.4 & 0.96 & 0.99 & 1.13 & All outcomes have effect $\delta_1$ \\ 
  0.4 & 0.2 & 0.2 & 0.81 & 0.82 & 0.99 & Effects as anticipated (power) \\ 
  0.4 & 0.0 & 0.0 & 0.76 & 0.30 & 0.87 & Two outcomes have no effect \\ 
  0.4 & -0.2 & -0.2 & 0.76 & 0.02 & 0.84 & Two outcomes are harmful \\ 
  0.0 & 0.0 & 0.0 & 0.02 & 0.02 & 0.96 & Global null (type-I error) \\ 
  0.3 & 0.3 & 0.3 & 0.78 & 0.90 & 1.07 & All have some effect $<\delta_1$ \\ 
  0.2 & 0.2 & 0.2 & 0.44 & 0.58 & 1.00 & All outcomes have effect $\delta_0$ \\ 
   \bottomrule
\end{tabular}
\caption{$R(\boldsymbol{\mu}= \mu_1, \mu_2, \mu_3)$ and expected sample size ratios for MO design and composite design, where $K=3, m=1, J=3, \boldsymbol{\delta}_\beta = (0.4, 0.2, 0.2)$.} 
\label{tab:true_mo}
\end{table}

\begin{figure}[htbp!]
    \centering
    \begin{subfigure}[htbp!]{0.45\textwidth}
    \includegraphics[width=\textwidth]{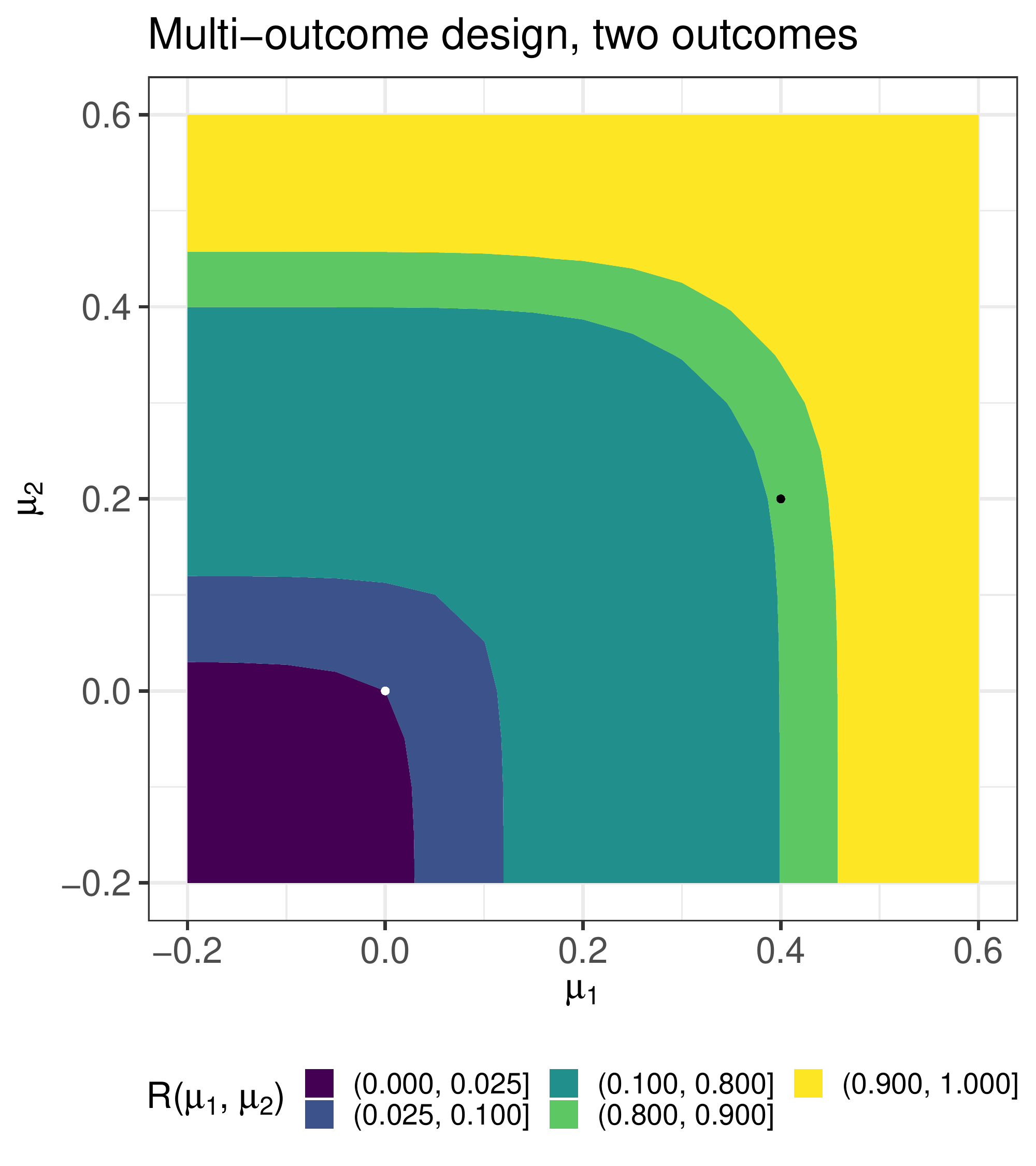}
    \caption{Multi-outcome multi-stage design realisation with $N=60, C=2.256490, \Delta=0$. Operating characteristics: $\alpha^*=0.025, 1-\beta^*=0.827$.}
    \label{fig:MO_2outcomes}
    \end{subfigure}
    ~
    \begin{subfigure}[htbp!]{0.45\textwidth}
    \includegraphics[width=\textwidth]{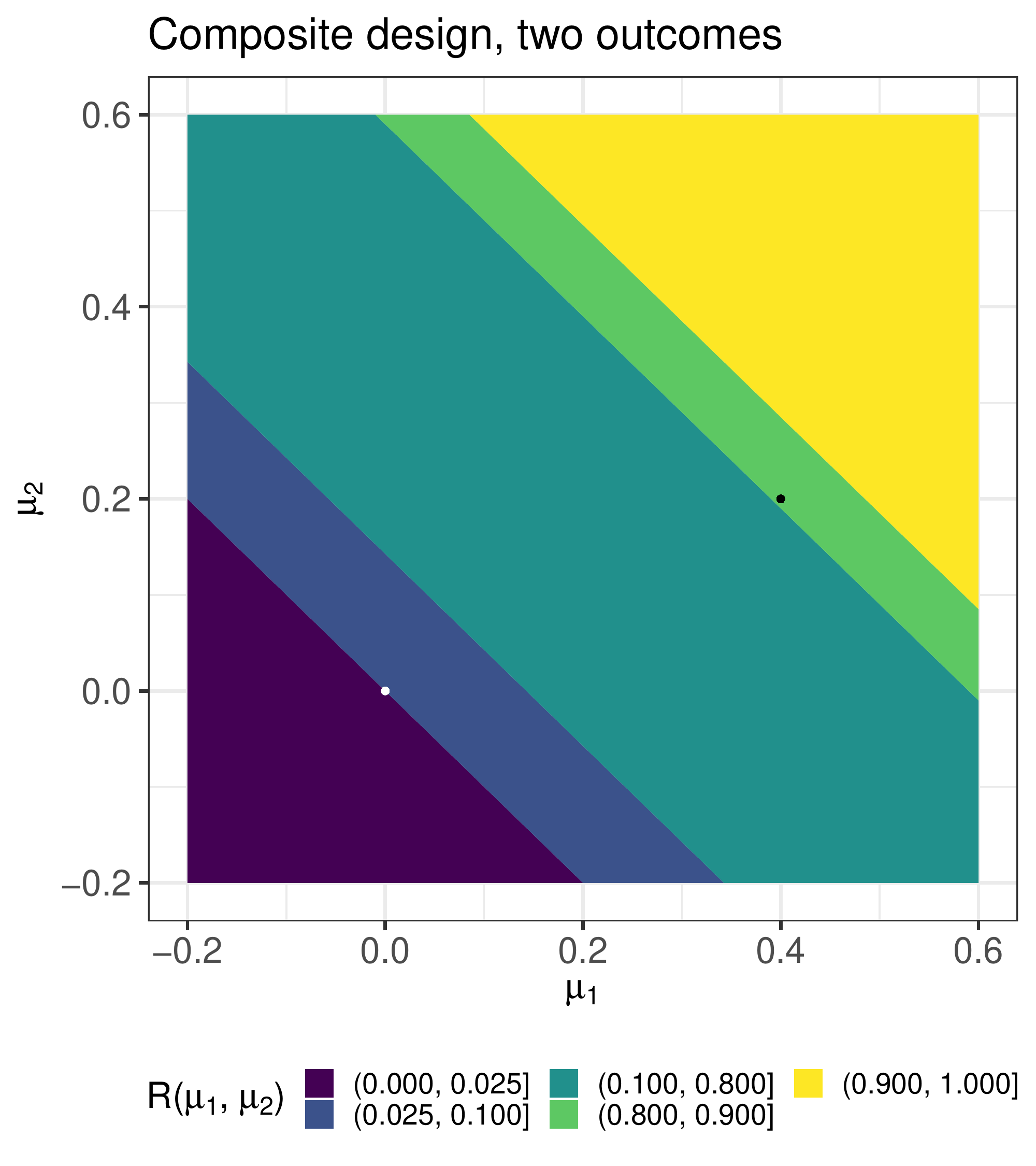}
    \caption{Composite multi-stage design realisation with $N=60, C=3.240066, \Delta=0$. Operating characteristics: $\alpha^*=0.025, 1-\beta^*=0.814$.}
    \label{fig:comp_2outcomes}
    \end{subfigure}  
    \caption{Probability of rejecting $H_0$ as true effect sizes vary. Powered for effect sizes $\boldsymbol{\delta}_\beta=(0.4, 0.2)$. Design parameters $K=2, m=1, J=3, \alpha=0.025$, $1-\beta=0.8,$ $\rho_{k_1k_2}=\rho=0.3, \, k_1  \neq k_2$, $\sigma^2_k=\sigma^2=1, \, \forall k$. White dot indicates global null $\boldsymbol{\mu} = \mathbf{0}$, black dot indicates point for which design is powered, $\boldsymbol{\mu} = \boldsymbol{\delta}_\beta$.}
    \label{fig:2outcome_rejection_regions}
\end{figure}

The idea that multi-outcome designs have rejection regions or spaces was introduced in Section~\ref{sec:MO_typeIerror_power}. We compare the different rejection regions of the multi-outcome multi-stage design with the composite design for $\{K=2, m=1, J=3\}$ and $\{K=3, m=2, J=3\}$. In Figure~\ref{fig:2outcome_rejection_regions}, we show $R(\mu_1, \mu_2)$, the probability of rejecting $H_0$ given true outcome effects $\mu_1, \mu_2$. The required type-I error-rate and power were $\alpha=0.025, 1-\beta=0.8$, with the designs powered for outcome effect sizes $\mu_1=0.4, \mu_2=0.2$. Suitable design realisations were obtained for $N=57$ (19 per stage) in the multi-outcome design and $N=60$ (20 per stage) for the composite design (as above). For this comparison, we wanted $N$ to be equal for the design realisations of both designs. Requiring $N=60$ for the multi-outcome design meant that power was increased, hence the power is greater than may be expected ($1-\beta^*=0.827$, black dot on Figure~\ref{fig:MO_2outcomes}). The black dots, representing the points for which the designs are powered, are do not lie exactly on a contour. Beyond the explanation for the increased power of the multi-outcome design above, this is due to the discrete nature of sample size: for these designs, sample size is increased until the required power is reached. The type-I error-rate is determined by the stopping boundary constant $C$, which may take any continuous value. Consequently, the white dots, indicating the point at which the type-I error-rate must be satisfied, both lie exactly on a contour.

The shapes of the regions largely reflect those in Figures~\ref{fig:rejection_region} and~\ref{fig:MO_preject_grid}: the group of regions within which the probability of rejection is low is approximately square for the multi-outcome design and triangular for the composite design. The reasoning remains the same: the multi-outcome design does not penalise a negative effect size, unlike the composite design. In general, the additive nature of the composite design plays a strong role in the differences between the regions.

\begin{figure}[htbp!]
    \centering
    \begin{subfigure}[htbp!]{0.48\textwidth}
    \includegraphics[width=\textwidth]{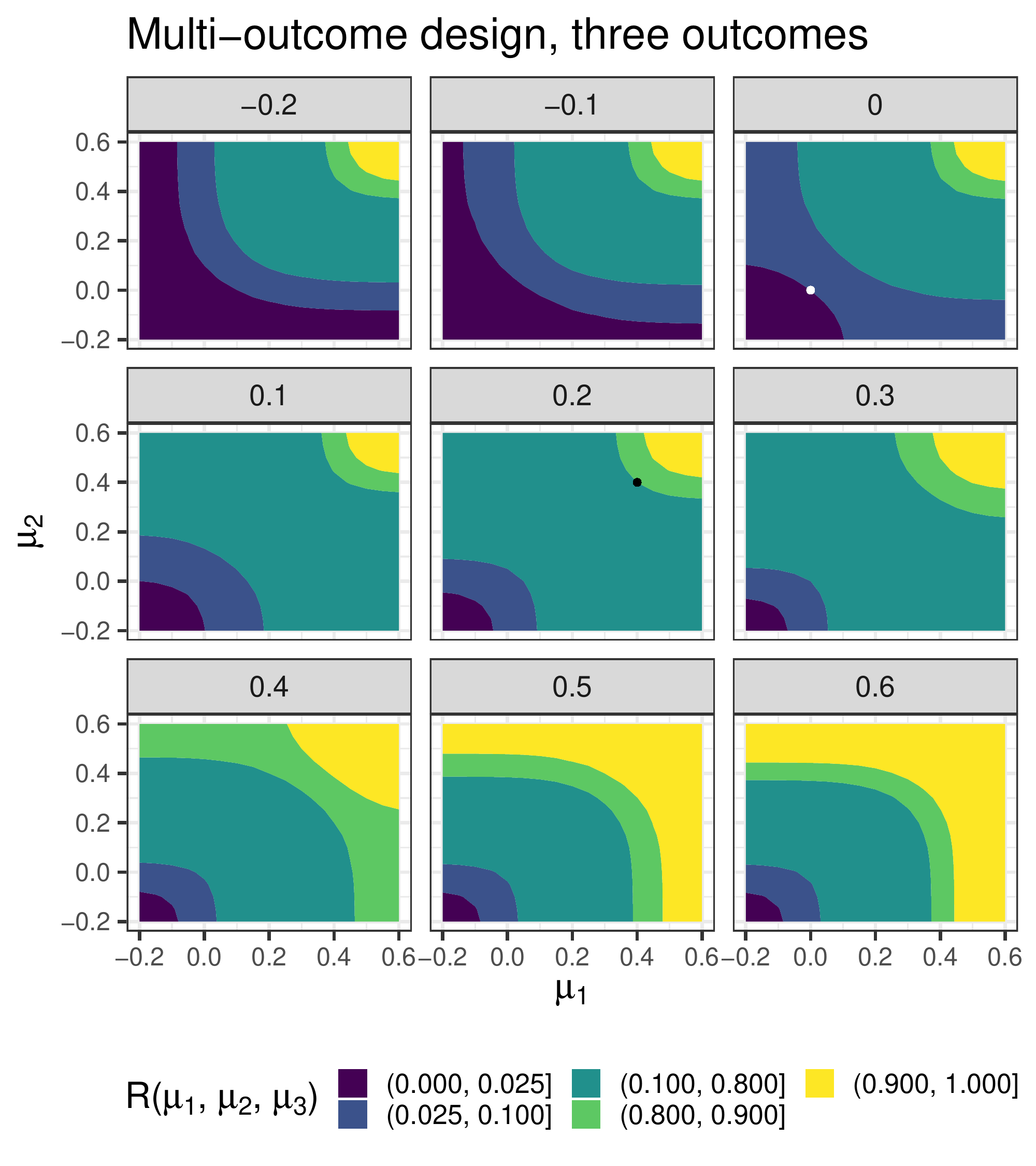}
    \caption{Multi-outcome multi-stage design realisation with $N=42, C=1.579395, \Delta=0$. Operating characteristics: $\alpha^*=0.025, 1-\beta^*=0.801$.}
    \label{fig:MO_3outcomes}
    \end{subfigure}
    ~
    \begin{subfigure}[htbp!]{0.48\textwidth}
    \includegraphics[width=\textwidth]{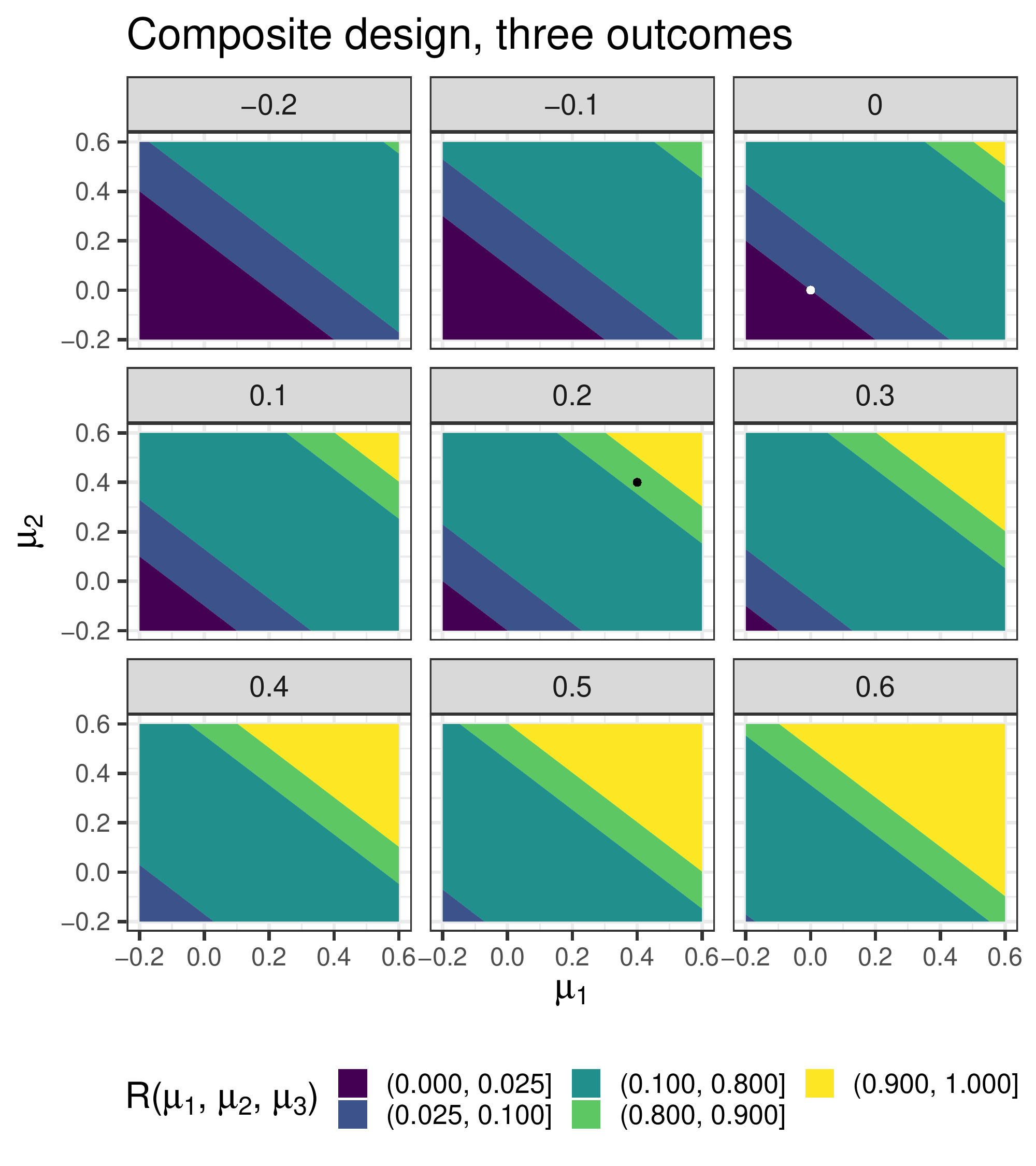}
    \caption{Composite multi-stage design realisation with $N=42, C=4.389363, \Delta=0$. Operating characteristics: $\alpha^*=0.025, 1-\beta^*=0.836$.}
    \label{fig:comp_3outcomes}
    \end{subfigure}  
    \caption{Probability of rejecting $H_0$ as true effect sizes vary. Powered for effect sizes $\boldsymbol{\delta}_\beta=(0.4, 0.4,  0.2)$. Design parameters $K=3, m=2, J=3, \alpha=0.025$, $1-\beta=0.8,$ $\rho_{k_1k_2}=\rho=0.3, \, k_1  \neq k_2$, $\sigma^2_k=\sigma^2=1, \, \forall k$. White dot indicates global null $\boldsymbol{\mu} = \mathbf{0}$, black dot indicates point for which design is powered, $\boldsymbol{\mu} = \boldsymbol{\delta}_\beta$. Each plot slice represents true effect size for $\mu_3$.}
    \label{fig:3outcome_rejection_regions}
\end{figure}

Figure~\ref{fig:3outcome_rejection_regions} shows rejection regions for three outcomes, powered to find two promising outcomes $\boldsymbol{\delta}_\beta = (0.4, 0.4, 0.2)$ and with three stages, that is, $\{K=3, m=2, J=3\}$. The sample size on the composite design was increased so that sample size was equal across both design realisations, $N=42$ (14 per stage). Again, the black dots indicating power do not lie exactly on a contour, in contrast with the white dots indicating type-I error-rate which do lie exactly on a contour. For  $\mu_3 \in \{0.5, 0.6\}$, rejection regions are similar to Figure~\ref{fig:2outcome_rejection_regions} (and again Figures~\ref{fig:rejection_region} and~\ref{fig:MO_preject_grid}). For non-positive values of $\mu_3$, the low probability of rejecting $H_0$ in the composite design can be seen, even when the remaining outcomes have considerable effect sizes. Conversely, for the corresponding plots on for multi-outcome design, non-positive values of $\mu_3$ have little effect on the size of the rejection regions. This again shows the nature of the difference between an additive and non-additive test statistic. As $\mu_3$ increases, the rejection regions of the composite design seem to shift linearly and without changing shape. However, in the multi-outcome design the regions corresponding to high probability of rejection change shape as $\mu_3$ increases, from a small square to a large inverted ``L'' shape. Conversely, the region corresponding to low probability of rejection changes shape in the opposite way. This is because when $\mu_3$ is low, it is little chance of this outcome contributing to a rejection of $H_0$. As $\mu_3$ increases closer to value for which the promising outcomes are powered, this probability increases. When $\mu_3$ is much greater than this, it is almost certain to contribute to the rejection of $H_0$ (by exceeding its stopping boundary). As such, only one of the two remaining outcomes $\mu_1, \mu_2$ are additionally required to show promise for $H_0$ to be rejected. Therefore $H_0$ is likely to be rejected when either one of $\mu_1, \mu_2$ shows promise. Furthermore, as rejection of $H_0$ is dependent on only (any) two outcomes showing an effect, there is little ``benefit'' from all three outcomes having large effect sizes. Indeed, for this design realisation, the probability of rejecting $H_0$ when any $\mu_k=\infty, k \in \{1, 2, 3\}$ is approximately 0.12, while this probability is necessarily equal to one for any composite design realisation.

\section{Methods: Drop the loser approach based on conditional probability, two-stage}
The multi-outcome multi-stage approach may be combined with a DtL-type component, that is, dropping an outcome (or outcomes) before the end of trial, with the aim of reducing ENM. This approach to reducing the number of measurements in a trial is an alternative to using separate stopping rules, as described in Section~\ref{sec:MOMS_design_1}.

Again, let $K$ be the number of outcomes and $m$ be the number of outcomes required to show promise in order to reject the null hypothesis. Fix the number of stages to be equal to 2. The shared final rejection boundary is given by $r$. If an outcome $k$ is not dropped at the interim analysis, that is, it is still being measured at the end of the trial, its test statistic $Z_{2k}$ will be compared against this final rejection boundary $r$. We again specify lower and greater anticipated treatment effects for each outcome, $\boldsymbol{\delta_0}=(\delta_{01}, \delta_{02}, \dots, \delta_{0K})$ and $\boldsymbol{\delta_1}=(\delta_{11}, \delta_{12}, \dots, \delta_{1K})$. Let the true outcome effects again be $\boldsymbol{\mu} = (\mu_1, \mu_2, \dots, \mu_K)$. 

The number of outcomes dropped at the interim analysis may be fixed in advance or determined by the interim data. In either case, some approach must be used to determine the ``losers'', the poorest-performing outcomes. The approach we have chosen is to use conditional power (CP)~\cite{Hamasaki2017, jennturn}. Here, we define $CP_k$ as the probability of outcome $k$ exceeding the final rejection boundary $r$, conditional on the data for outcome $k$ observed so far and an anticipated outcome effect $\delta_{1k}$. For a general number of stages $j, j=1, \dots, J,$ the conditional power of outcome $k$ at stage $j$ is then $CP_{jk}(\delta_{1k}) = \text{P}(Z_{Jk} > r \vert Z_{jk}, \delta_{1k})$. The calculation for the conditional power of outcome $k$ at the single interim analysis, given current data and anticipated outcome effect $\delta_{1k}$ is 

\begin{equation}
\label{eq:jennturn}
CP_{k} (\delta_{1k}) =  \Phi \left( \frac{Z_{k}\sqrt{\mathcal{I}_1} - r\sqrt{\mathcal{I}_2} + (\mathcal{I}_2 - \mathcal{I}_1)\delta_{1k}}{\sqrt{(\mathcal{I}_2 - \mathcal{I}_1)}} \right).
\end{equation}

Equation~(\ref{eq:jennturn}) is merely a special case of the equation for a general number of stages $j$ provided by Jennison and Turnbull~\cite{jennturn}. As in Section~\ref{sec:MOMS_design_1}, using a single shared boundary avoids a $K$-dimensional optimisation problem with infinite solutions.

\subsection{Conditional power-based stopping (and dropping) boundaries}\label{sec:MO_CP}
CP is used in this multi-outcome multi-stage design, rather than comparing test statistics to boundaries directly as may be expected in the multi-arm setting. However, multi-arm trials are generally used to evaluate if any treatment has some single effect size of interest. In contrast, outcomes may have different anticipated effect sizes. As such, absolute values of test statistics may not give an accurate indication of the relative interim performance of outcomes. For example, among two interim test statistics, one test statistic ($Z_{11}$ say) may be lower than another ($Z_{12}$) while being closer to its anticipated standardised effect size, i.e. $(\delta_{11}/\sigma_1) - Z_{11} < (\delta_{12}/\sigma_2) - Z_{12} , \quad (\delta_{11}/\sigma_1) < Z_{11}, \, (\delta_{12}/\sigma_2) < Z_{12}$. In this case, the outcome with the lower test statistic may be the outcome that is more likely to exceed the final rejection boundary, and should not necessarily be the outcome that is dropped.



We specify lower and upper interim stopping boundaries in terms of some conditional probabilities $CP_L$ and $CP_U$. Our approach to dropping outcomes and to stopping the trial are as follows: if the CP of the test statistic of some outcome $k$ is less than $CP_L$ at the interim, that is, $CP_k(\delta_{1k}) < CP_L$, it is dropped from the trial and not measured nor evaluated at the final stage. If $K-m+1$ or more outcomes are dropped, the trial ends early for a no-go decision. If the CPs of the test statistics of $m$ or more outcomes are greater than $CP_U$ at the interim, that is, if $\sum_{k=1}^K \mathbb{I} \left(CP_k(\delta_{1k}) > CP_U \right) \geq m$, the trial ends early for a go decision. If the trial does not end early, it proceeds to a second stage. The number of outcomes retained for stage 2 is

\[
K_2 = min\left( K_{max}, \; \sum\limits_{k=1}^K \mathbb{I} \left(CP_k(\delta_{1k}) > CP_L \right) \right)
\]

for some fixed $K_{max} < K$. The value of $K_{max}$ determines the maximum possible number of outcomes that may be measured in stage 2, and thus also determines the maximum number of outcome measurements obtained in this design approach.

The null hypothesis is unchanged compared to the previous multi-outcome multi-stage approach, given by Equation (\ref{eq:H0_moms}). The null hypothesis is rejected if either the trial continues to stage 2 and at least $m$ retained outcomes exceed the final stopping boundary $r$, or if at least $m$ $CP$ values exceed $CP_U$ at the interim, that is if

\[
\left( \sum\limits_{k=1}^K CP_k(\delta_{1k}) > CP_L \cap \mathbb{I} (Z_{2k} > r) \right) \geq m \qquad \text{ and } \qquad \sum\limits_{k=1}^K \mathbb{I} (CP_k(\delta_{1k}) > CP_U) < m 
\]
\[
\text{or} \qquad \sum\limits_{k=1}^K \mathbb{I} (CP_k(\delta_{1k}) > CP_U) \geq m.
\]

As with the first proposed multi-outcome approach, we define the probability of rejecting the null hypothesis for outcome effects $\boldsymbol{\mu}$, but for this approach the design parameters are $K, K_{max},$ $m, CP_L,$ $CP_U$. We define type-I error-rate as the probability of rejecting the null hypothesis under the global null, $\alpha^* = R(\boldsymbol{\mu} = \mathbf{0} \vert K, K_{max}, m, CP_L, CP_U)$ and the power as the probability of rejecting the null hypothesis under the LFC, $1-\beta^* = R(\boldsymbol{\mu} = \boldsymbol{\delta}_\beta \vert  K, K_{max}, m, CP_L, CP_U)$ similar to Section~\ref{sec:MOMS_design_1}.

\subsection{Design search}
To search for designs, sets of $2K$ test statistics are simulated under the global null hypothesis $\boldsymbol{\mu}=\mathbf{0}$. A search is undertaken to find a design that fulfils the required type-I error-rate $\alpha$ and power $1-\beta$. The interim boundaries $CP_L$,  $CP_U$ and anticipated effects $\boldsymbol{\delta_0}, \boldsymbol{\delta_1}$ are fixed and specified in advance. The operating characteristics of a trial therefore depend on the final rejection boundary $r$ and the per-stage sample size $n$. A shared rejection boundary $r$ is found that minimises $(\alpha - \alpha^*)^2$ for some initial per-stage $n$. Using these boundaries and $n$, $1-\beta^*$ is obtained. If $1-\beta^*$ is less than the required power $1-\beta$, then the process of finding $r$ and power is repeated with increased $n$. Conversely if $1-\beta^*$ is greater than the required power $1-\beta$, the process is repeated with decreased $n$. Thus the per-stage sample size $n$ is altered to find the smallest value that satisfies the required power. Some $nsims$ number of trials are simulated as in Section~\ref{sec:mo_design_search}. The rest of the design search is described in Algorithms~\ref{algo:MO_findCP}, \ref{algo:MO_pRejectDTL} and~\ref{algo:MO_full_search_DtL}.

\begin{algorithm}[htbp!]
\SetAlgoLined
    numerator $\leftarrow$ TSrow$\sqrt{\mathcal{I}_1} - r \sqrt{\mathcal{I}_2} + (\mathcal{I}_2 - \mathcal{I}_1) \boldsymbol{\delta}_1$ \;
    denominator $\leftarrow \sqrt{(\mathcal{I}_2 - \mathcal{I}_1)}$ \;
    cp $\leftarrow$ call normalCDF(numerator/denominator)\;
\caption{findCPs: find conditional power at stage 1, for a vector of outcomes. Input:  TSrow (one row of $K$ simulated interim test statistics), $\mathcal{I}_1, \mathcal{I}_2, \boldsymbol{\delta}_1, r$.}
\label{algo:MO_findCP}
\end{algorithm}

\begin{algorithm}[htbp!]
\footnotesize
\SetAlgoLined
$\mathcal{I}_1 \leftarrow$ call findInformation(n.per.stage, $\boldsymbol{\sigma^2}, K$)\;
$\mathcal{I}_2 \leftarrow$ call findInformation(2*n.per.stage, $\boldsymbol{\sigma^2}, K$)\;
\If{type1.err.or.power==``power''}{
    $\boldsymbol{\delta}_\beta \leftarrow$ call findDeltaBeta($\boldsymbol{\delta}_0, \boldsymbol{\delta}_1$)\;
    $\boldsymbol{\tau} \leftarrow$ call findEffects($\boldsymbol{\delta}_\beta, \mathcal{I}_1, \mathcal{I}_2$)\;
    TS $\leftarrow$ call addEffectsToTS(TS, $\boldsymbol{\tau}$)\;
}
TS.stage1 $\leftarrow$ TS[, 1:$K$]\;
TS.stage2 $\leftarrow$ TS[, ($K+1$):$2K$]\;
nsims $\leftarrow$ nrow(TS)\;
\For{$i$ in 1 to nsims}{
    CPs[$i$, ] $\leftarrow$ call findCPs(TS.stage1[$i$, ], $\mathcal{I}_1, \mathcal{I}_2, r, \boldsymbol{\delta}_1$)\;
    \eIf{sum(CPs[$i$, ] $<$ $CP_L$) $\geq K-m+1$}{
    no.go.decision.stage1[$i$] $\leftarrow$ 1 \;
    go.decision.stage1[$i$] $\leftarrow$ 0 \;
    }{\eIf{sum(CPs[$i$, ] $> CP_U$) $\geq m$}{
        no.go.decision.stage1[$i$] $\leftarrow$ 0 \;
        go.decision.stage1[$i$] $\leftarrow$ 1 \;}{
            no.go.decision.stage1[$i$] $\leftarrow$ 0 \;
            go.decision.stage1[$i$] $\leftarrow$ 0 \;}
    }
}
stop.early $\leftarrow$ no.go.decision.stage1 + go.decision.stage1 \;
continue $\leftarrow$ !stop.early \;
TS.continue $\leftarrow$ call subsetToContinuingTrials(TS.stage2, continue)\;
CPs.continue $\leftarrow$ call subsetToContinuingTrials(CPs, continue)\;
nrows.continue $\leftarrow$ sum(continue)\;
\For{$i$ in 1 to nrows.continue}{
    CPs.ranked $\leftarrow$ call rankCPs(CPs.continue[$i$, ])\;
    retained.outcomes[$i$] $\leftarrow$ call retainGreatestCPs(CPs.ranked, $K_{max}, CP_L$)\;
    retained.TSs $\leftarrow$ call subsetToRetainedTSs(TS.continue[$i$, ], retained.outcomes[$i$])\;
    \eIf{sum(retained.TSs $> r$) $\geq m$}{
        go.decision.stage2[$i$] $\leftarrow$ 1 \;
        }{
        go.decision.stage2[$i$] $\leftarrow$ 0 \;
    }
}
no.measurements.stage2 $\leftarrow$ sum(retained.outcomes) \;
prob.reject $\leftarrow$ go.decision.stage1 + go.decision.stage2 \;
PET $\leftarrow$ sum(stop.early)/nsims \;
ENM $\leftarrow$ $K$ + no.measurements.stage2/nsims \;
\If{type1.err.or.power==``typeIerror''}{
    minimise.prob $\leftarrow ( \text{prob.reject} - \alpha ) ^2$ \;
}
\caption{Function pRejectDTL: for finding type-I error-rate or power, expected number of stages and ENM for DtL approach. Input: $r, K, K_{max}$ n.per.stage (current $n$ per stage), $m, \boldsymbol{\delta}_0$, $\boldsymbol{\delta}_1, CP_L$, $CP_U, \boldsymbol{\sigma}^2$, $ \alpha,$ $TS \text{ (matrix of test statistics)}$, type1err.or.power (whether finding type-I error-rate or power)}
\label{algo:MO_pRejectDTL}
\end{algorithm}

\begin{algorithm}[htbp!]
\SetAlgoLined
n.vec $\leftarrow$ nmin:nmax \;
$a$ $\leftarrow$ 1\;
$b$ $\leftarrow$ length(n.vec)\;
$d$ $\leftarrow$ ceiling((b-a)/2)\;
\While{$b-a>1$}{
    r.current $\leftarrow$ call optimise(pRejectDTL, typeIerror.or.power=``typeIerror'', n.per.stage=n.vec[$d$], $\dots$)\;
    pow $\leftarrow$ call pRejectDTL(r.current, typeIerror.or.power=``power'', n.per.stage=n.vec[$d$], $\dots$)\;
    \eIf{pow $<$ power}{
    $a$ $\leftarrow$ $d$\;
    $d$ $\leftarrow$ ceiling($a+(b-a)/2$)\;
    }{
        $b$ $\leftarrow$ $d$\;
        $d$ $\leftarrow$ ceiling($a+(b-a)/2$)\;
    }
}
n.final $\leftarrow$ n.vec[$d$]\;
r.final $\leftarrow$ call optimise(pRejectDTL(n.per.stage=n.final, typeIerror.or.power=``typeIerror''))\;
typeIerr.output $\leftarrow$ call pRejectDTL(r.final, typeIerror.or.power=``typeIerror'', n.per.stage=n.final, $\dots$)\;
power.output $\leftarrow$ call pRejectDTL(r.final, typeIerror.or.power=``power'', n.per.stage=n.final, $\dots$)\;
$\alpha^*$ $\leftarrow$ call selectPReject(typeIerr.output)\;
pow $\leftarrow$ call selectPReject(power.output)\;
$N$ $\leftarrow$ 2*n.final\;
$ESS_0$ $\leftarrow$ call selectPET(typeIerr.output)*n.final + (1-selectPET(typeIerr.output))*$N$\;
$ESS_1$ $\leftarrow$ call selectPET(power.output)*n.final + (1-selectPET(power.output))*$N$\;
$ENM_0$ $\leftarrow$ call selectENM(typeIerr.output)*n.final\;
$ENM_1$ $\leftarrow$ call selectENM(power.output)*n.final\;
\caption{findDtLDesign: find DtL design realisation that satisfies required type-I error-rate and power. Input: TS, nmin, nmax, $m, K, K_{max}$, $\alpha, 1-\beta$, $\mathcal{I}_1, \mathcal{I}_2, \boldsymbol{\delta}_0, \boldsymbol{\delta}_1,$ $CP_L, CP_U, \boldsymbol{\sigma^2}$.}
\label{algo:MO_full_search_DtL}
\end{algorithm}

\section{Results: drop the loser approach based on conditional probability, 2-stage}
\subsection{Varying correlation}
The multiple outcome DtL approach was compared to a multiple outcome single-stage approach in terms of ESS and ENM ratios (denoted $ESS_{DtL}/ESS_{single}$ and $ENM_{DtL}/ENM_{single}$)the under LFC as correlation varied $(\rho = \{0, 0.1, \dots, 0.8 \})$. Design realisations were found for $\{K, m \}= \{2, 1\}, \{6, 1\}, \{6, 3\}$ and $K_{max} = \{K-1, K/2\}$ (see Table~\ref{tab:DtL_design_params}). Other design parameters were as the previous approach (Section~\ref{sec:mo_comparing_des_part1}): $\alpha=0.025, 1-\beta=0.8, \delta_{01}=\delta_{02}= \dots = \delta_{0K}=\delta_0=0.2, \delta_{11}=\delta_{12}= \dots = \delta_{1K}=\delta_1=0.4, \sigma^2_k=1, \, \forall k$. The lower and upper conditional power thresholds were set to $CP_L=0.3$ and $CP_U=0.95$ respectively. For some previous uses of conditional power, the maximum lower threshold for CP was set equal to the response rate for which the trial was powered~\cite{law,law2020stochastically}. Here, there is no such obvious association to be made between CP threshold, a probability and effect size, a continuous value. In the absence of suggestions in the literature, the thresholds $CP_L=0.3, CP_U=0.95$ were chosen. The admissible single-stage designs of Law et al.~\cite{law} often have similar thresholds, though we acknowledge the difference in the design approaches. At the trial planning stage, we recommend undertaking a sensitivity analysis of the interim thresholds, to more fully understand how the choice may affect a particular set of design parameters. Note that setting $CP_L=0$ is equivalent to permitting early stopping for a go decision only, while setting $CP_U=1$ is equivalent to permitting early stopping for a no go decision only.

\begin{table}[htbp!]
    \centering
    \begin{tabular}{ccl}
    \toprule
     $K$  & $m$ & $K_{max}$ \\
     \midrule
     2 & 1 & 1 ($K_{max}= K-1, K/2$)\\
     6 & 1 & 3 ($K_{max}= K/2$)\\
     6 & 1 & 5 ($K_{max}= K-1$)\\
     6 & 3 & 3 ($K_{max}= K/2$)\\
     6 & 3 & 5 ($K_{max}= K-1$)\\
    \bottomrule
    \end{tabular}
    \caption{Sets of design parameters $\{K, m, K_{max}\}$ used in comparison of proposed DtL design and single-stage design.}
    \label{tab:DtL_design_params}
\end{table}

\begin{figure}[htbp!]
\centering
\includegraphics[width=0.8\textwidth]{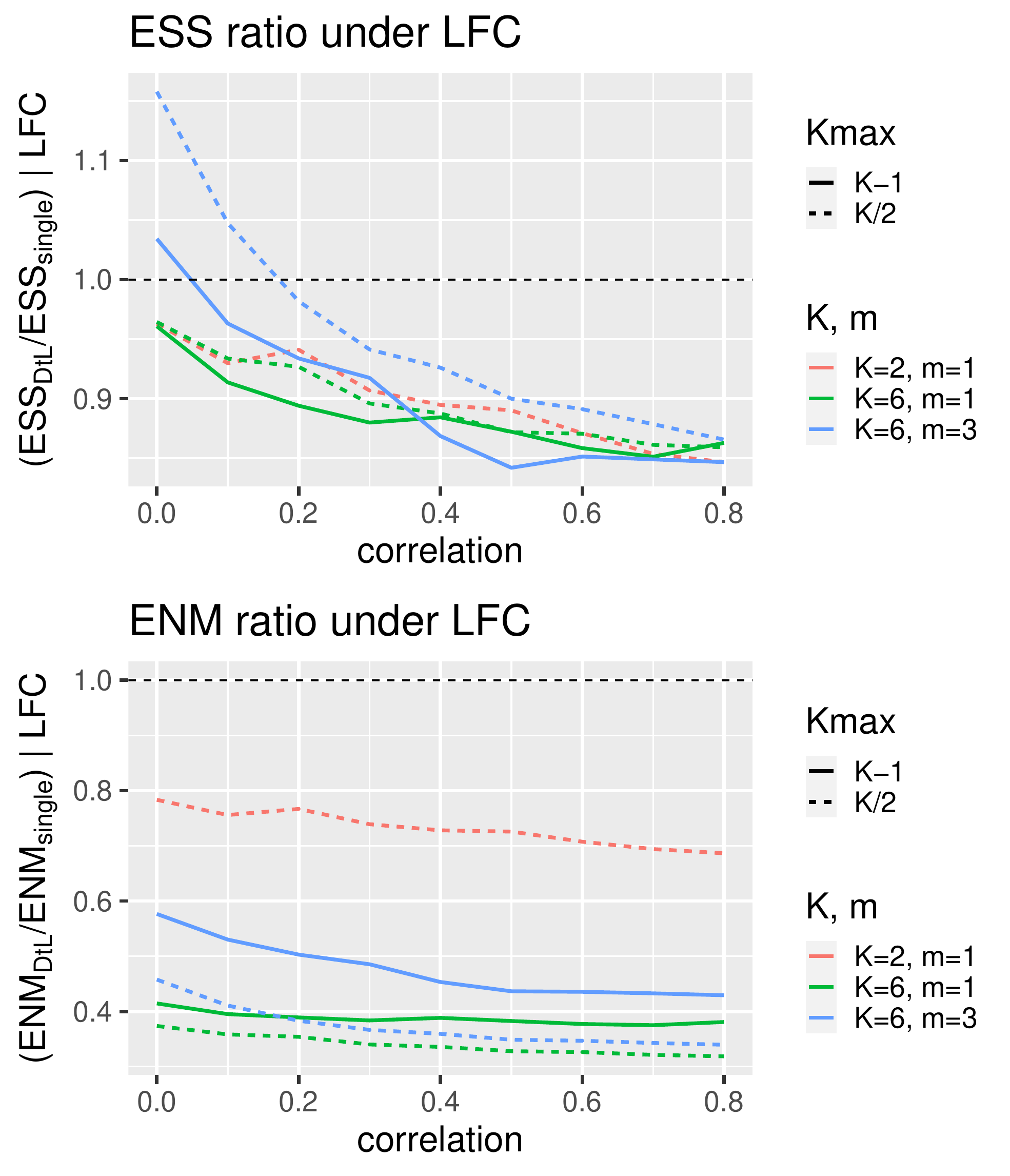}
\caption{Changes in $ESS_{DtL}/ESS_{single}$ and $ENM_{DtL}/ENM_{single}$ for various designs as correlation $\rho$ is varied. Note: for $\{K=2, m=1\}, K-1=K/2$.}
\label{fig:dtl_cor}
\end{figure}

The results are shown in Figure~\ref{fig:dtl_cor}. Values below 1 indicate superiority of the proposed DtL approach over the single stage approach. Similarly to the previous results, ESS ratio decreases as correlation $\rho$ increases. However, in this comparison, the ESS ratio is less than 1 in almost all cases, and in all but one case when $\rho>0$, though the ESS ratio is generally closer to 1 compared to the results in Figure~\ref{fig:MO_cor}. ESS ratio appears to be greater when $m>1$, though this difference seems to decrease as $\rho$ increases. The ENM ratio also decreases as $\rho$ increases. The ENM ratio is less than 1 in every case, meaning that fewer measurements are expected over both stages of the DtL design than in the single stage design. The ENM ratio is considerably greater under $\{K=2, m=1\}$ compared to the other combinations of $\{K, m\}$ examined. Using $K_{max}=K/2$ resulted in a lower ENM ratio than using $K_{max}=K-1$. This may be expected, as fewer outcomes are permitted to be retained for the second stage.

\subsection{Varying true outcome effects}
\subsubsection{Two outcomes}
\begin{figure}[htbp!]
\centering
\includegraphics[width=0.75\textwidth]{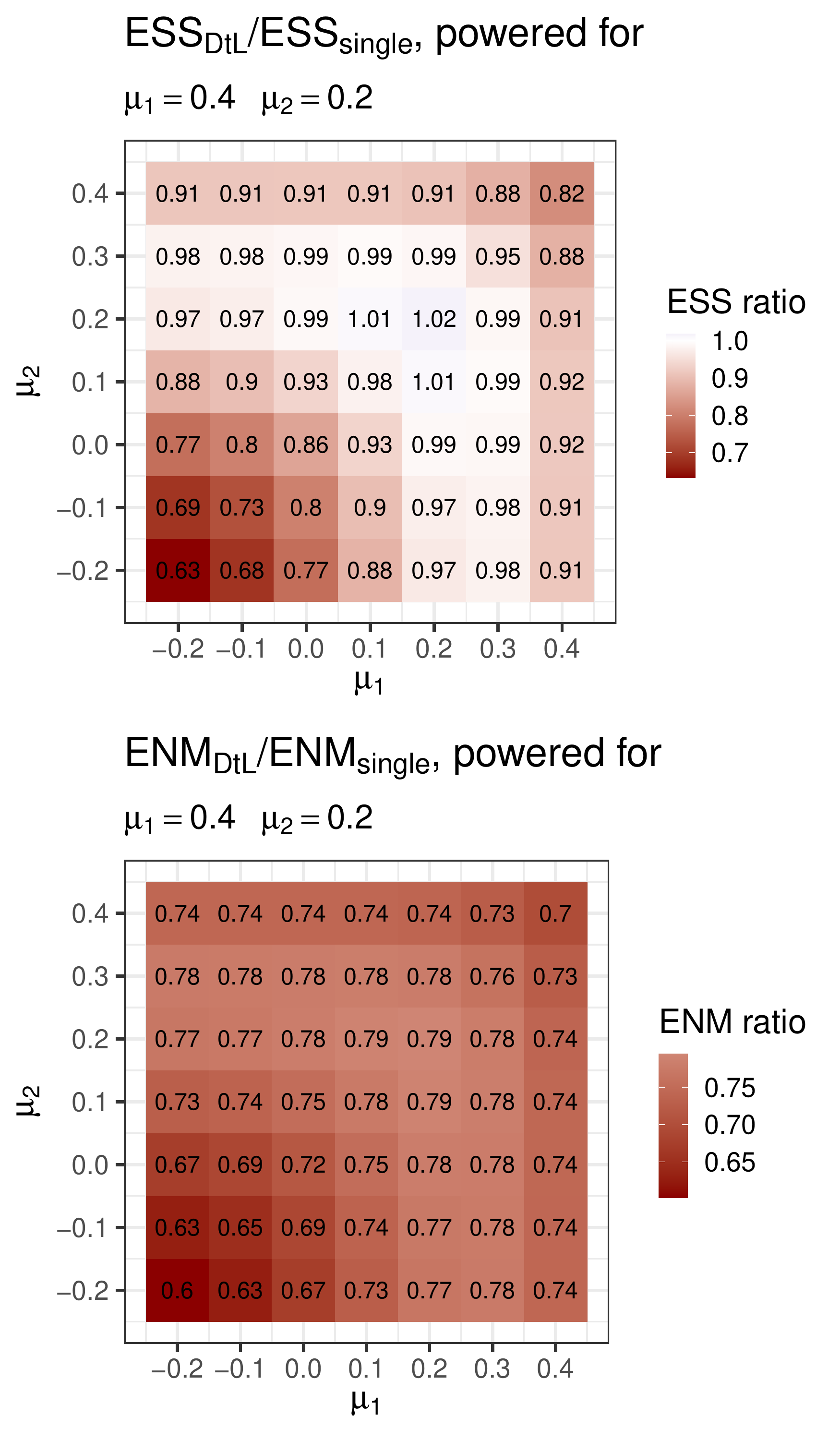}
\caption{Changes in $ESS_{DtL}/ESS_{single}$ and $ENM_{DtL}/ENM_{single}$ for fixed design with $\{K=2, K_{max}=1, m=1\}$ and design is powered for outcome effects $\boldsymbol{\delta}_\beta = (0.4, 0.2)$.}
\label{fig:dtl_ess_enm}
\end{figure}

The changes in ESS and ENM ratio for single design realisations as true outcome effects vary over $\mu_1, \mu_2 \in \{-0.2, -0.1, \dots, 0.4\}$ are shown in Figure~\ref{fig:dtl_ess_enm}. Design parameters are $\{K=2, K_{max}=1, m=1\}, \boldsymbol{\delta}_\beta = (0.4, 0.2)$. The design realisations are $\{r=2.273714, N=64\}$ for the DtL design and $\{r=2.221584, N=56\}$ for the single-stage design.

As in Figure~\ref{fig:dtl_cor}, ESS ratio is generally less than 1, with ESS being lower in the single stage design in just three out of 49 cases. This occurs when both $\mu_1, \mu_2 \approx \delta_{\beta 2} = 0.2$. The greatest disparity in ESS is when $\mu_1 = \mu_2 = -0.2$, the minimum effect size examined. When the true effect sizes are low, or even harmful, the conditional power will be low and the possibility of early stopping increases. The ENM ratio shows similar results, with the lowest values (and greatest benefit of the DtL design) observed when the true outcome effects are at their lowest with either trial ending or dropping an outcome at the interim. The ENM ratio is less than 1 in all cases.

\begin{figure}[htbp!]
\centering
\includegraphics[width=0.7\textwidth]{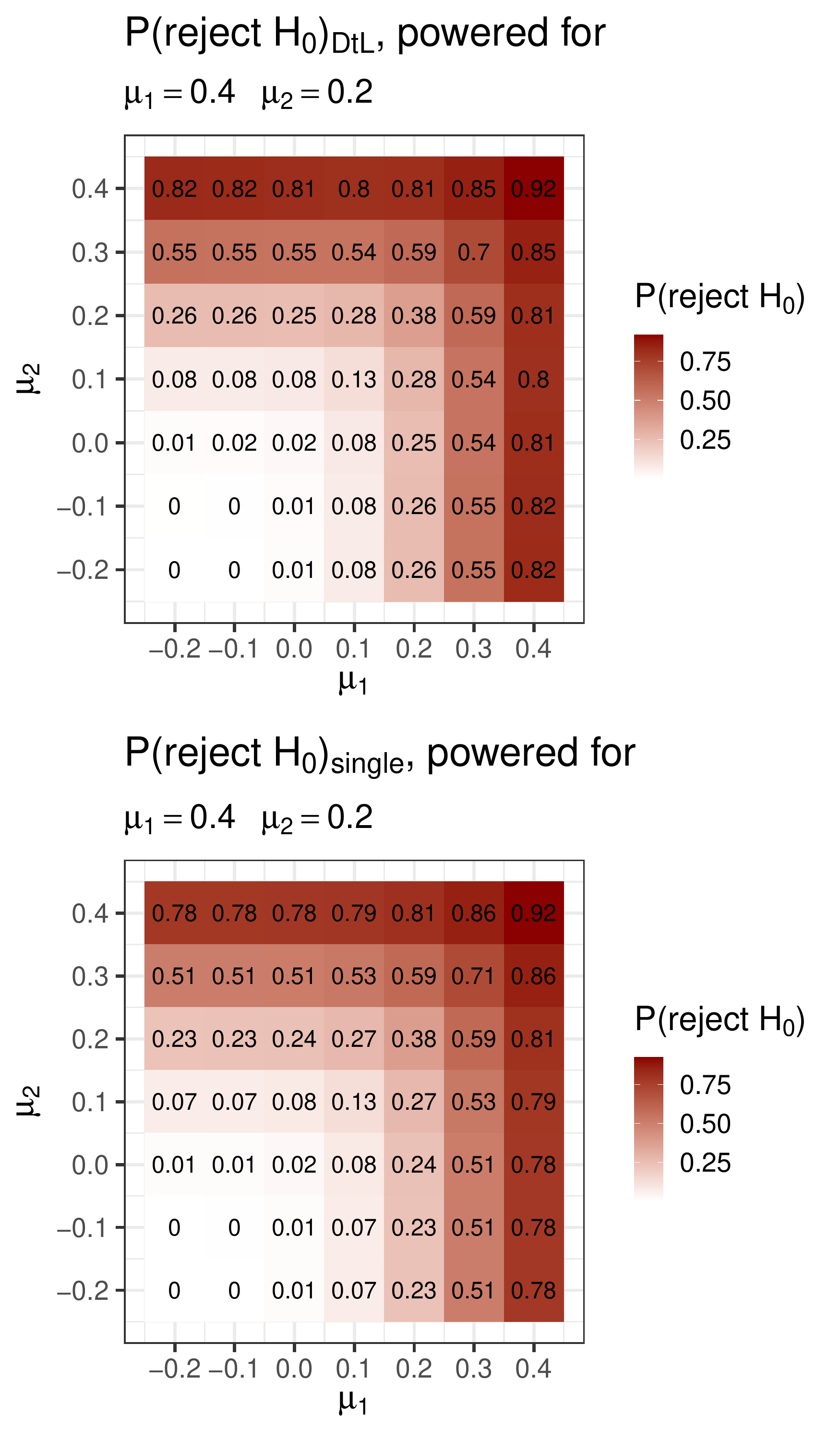}
\caption{Changes in the probability of rejecting $H_0$ for the DtL and single-stage designs with $\{K=2, K_{max}=1, m=1\}$ and design is powered for outcome effects $\boldsymbol{\delta}_\beta=(0.4, 0.2)$.}
\label{fig:dtl_preject}
\end{figure}

For the same design realisations, the probability of rejecting $H_0$ under each approach is shown for $\mu_1, \mu_2 \in \{-0.2, -0.1, \dots, 0.4\}$ in Figure~\ref{fig:dtl_preject}. Both approaches show increases as one or both effect sizes increase. For all cases such that one outcome has the anticipated effect size 0.4 while the other has an effect size of 0.1 or lower, the DtL design reports a probability of rejecting $H_0$ slightly greater than nominal $[0.80, 0.82]$, possibly due to a slightly increased probability of dropping the poorly-performing outcome over the better-performing outcome compared to having effect sizes of $\boldsymbol{\mu}=(0.4, 0.2)$. For the same cases, the single stage design reports a probability slightly lower than nominal $[0.78, 0.79]$, possibly due to a slightly decreased probability of of rejecting $H_0$ due to the poorly-performing outcome.

\subsubsection{Three outcomes}
In the case of $K=3, K_{max}=1, m=1$, probability of rejecting $H_0$, ESS ratio and ENM ratio are examined for a selection of true effect sizes $\{\mu_1, \mu_2, \mu_3 \}$ in Table~\ref{tab:dtl_table_K3}. The design realisations are $\{r=2.435647, N=72\}$ for the DtL design and $\{r=2.380403, N=59\}$ for the single-stage design. The results are in agreement with the $K=2$ case: probability of rejecting $H_0$ is similar for both approaches for most featured cases and slightly lower for the single stage approach when one outcome has true effect as anticipated and the remaining outcomes have zero or harmful true effects. ESS ratio is greater than 1, i.e., favouring the single stage design, only when all outcomes have effects equal to $\delta_0$. Again the ENM ratio is less than 1 in all cases.

\begin{table}[htbp!]
\centering
\begingroup\footnotesize
\begin{tabular}{rrrrrrrl}
  \toprule
$\mu_1$ & $\mu_2$ & $\mu_3$ & p(rej. $H_0)_{DtL}$ & p(rej. $H_0)_{SS}$ & $ESS_{DtL/SS}$ & $ENM_{DtL/SS}$ & Description \\ 
  \midrule
0.4 & 0.4 & 0.4 & 0.95 & 0.96 & 0.80 & 0.47 & All outcomes have effect $\delta_1$ \\ 
  0.4 & 0.2 & 0.2 & 0.81 & 0.80 & 0.95 & 0.52 & Effects as anticipated (power) \\ 
  0.4 & 0.0 & 0.0 & 0.82 & 0.76 & 0.97 & 0.53 & One outcome has no effect \\ 
  0.4 & -0.2 & -0.2 & 0.83 & 0.76 & 0.97 & 0.53 & One outcome is harmful \\ 
  0.0 & 0.0 & 0.0 & 0.02 & 0.02 & 0.96 & 0.52 & Global null (type I error) \\ 
  0.3 & 0.3 & 0.3 & 0.77 & 0.77 & 0.97 & 0.53 & All have some effect $<\delta_1$ \\ 
  0.2 & 0.2 & 0.2 & 0.43 & 0.43 & 1.09 & 0.57 & All outcomes have effect $\delta_0$ \\ 
   \bottomrule
\end{tabular}
\endgroup
\caption{p(reject $H_0 \vert \text{ true effects } \boldsymbol{\mu}$), expected sample size ratios and expected number of measurements ratios for drop the loser design and single stage design, where $\{K=3, K_{max}=1, m=1\}$ and design is powered for outcome effects $\boldsymbol{\delta}_\beta=(0.4, 0.2, 0.2)$. $ESS_{DtL/SS}$: ESS ratio. $ENM_{DtL/SS}$: ENM ratio.} 
\label{tab:dtl_table_K3}
\end{table}

\section{Discussion}

Measurement of multiple outcomes is typical in clinical trials, for a number of reasons discussed above. Amongst clinical trial designs where multiple key outcomes are measured, designs are generally powered to identify when either at least one outcome shows promise or all measured outcomes show promise. This work generalises this concept by presenting two multi-outcome designs, both of which are powered to find when at least some specified number of outcomes shows promise. One design permits any number of stages, while the other is a two-stage design that permits dropping outcomes that are performing poorly. Our main goal was to improve on existing designs by creating designs that meet the needs of investigators in ways that existing trials do not. In particular, both designs offer a generalised framework in terms of seeking a specified number of efficacious outcomes, and this framework is novel in a multi-stage setting.


The first approach, a multi-outcome multi-stage design, was compared to a multi-stage design with a single composite outcome. As outcome correlation increases, the ESS and ENM of the proposed approach decrease relative to the composite approach, and were superior in all tested cases with correlation $\rho \geq 0.5$. When different true outcome effects are examined, ESS and ENM are generally lower for the proposed design, and are only greater than the composite design when more outcomes are efficacious than anticipated. The probability of rejecting the null hypothesis is more robust under the proposed approach, remaining close to nominal power when some true outcome effects are lower than anticipated while this probability decreases under the composite approach. Furthermore, rejecting the null hypothesis when no outcomes have the desired effect size is less likely using the proposed approach.

The second approach, a multi-outcome, two-stage DtL design, was compared to a single-stage design. Again the ESS and ENM of the proposed approach decrease compared to the existing approach as correlation increases. ENM was superior in the proposed approach for all cases examined, while ESS was superior in 42 out of 45 cases. Furthermore, in greater than 50\% of cases, the ENM was reduced by at least half. When different true outcome effects were examined, ENM was reduced under the proposed DtL design compared to the single stage design in all cases, while ESS was reduced in 46 out of 49 cases. The probability of rejecting the null hypothesis when one outcome was as efficacious as anticipated while other outcomes had lower effect sizes than anticipated was similar for both approaches. However, the rejection probability was slightly greater than the required power for the DtL approach and slightly lower for the single stage approach.

\subsection{Limitations}
The proposed multi-outcome designs have limitations regarding their generality. Both rely on continuous outcomes, rather allowing other outcome types, such as binary or ordinal, either on their own or in combination. The designs are single-arm, rather than two-arm. There is a single final rejection boundary shared between all outcomes, rather than permitting different boundaries for each outcome. The second design permits only two stages, in contrast to the multi-arm DtL design, where a general number of stages are accommodated.

\subsection{Recommendations}\label{sec:Discussion_recommendations}
The proposed multi-outcome designs would be of value for any investigator who seeks to conduct a multi-outcome trial that is powered to identify when a specified number of the (continuous) outcomes show promise. In particular, the multi-outcome multi-stage design shows improvements in ESS and ENM compared to a multi-stage composite design when correlation between outcomes is high ($\rho \geq 0.5$). The multi-outcome DtL design also improves ESS and ENM when correlation is high, compared to the multi-outcome single-stage design, and so both designs are recommended when outcome correlation is (or is anticipated to be) high. The proposed designs can also be recommended when creating a composite outcome is not appropriate. Both proposed designs seek to reduce ENM, the multi-outcome multi-stage design by providing multiple interim analyses at which points the trial may stop for either a go or no go decision, the multi-outcome DtL design by allowing measurement of poorly performing outcomes to cease as well as allowing the trial to stop at a single interim analysis. As such, we recommend these designs when the cost of outcome measurement is high. Both designs showed less sensitivity to their comparators when the true effect sizes deviated from the anticipated effect sizes. As such, like the proposed two-arm outcome binary design above, we recommend the proposed multi-outcome designs when there is uncertainty regarding the anticipated effect sizes.

Both multi-outcome design approaches find design realisations using simulation, and such simply report a single design realisation, again by calling a single function in \textit{R}~\cite{R}. The multi-outcome multi-stage design finds a single design and reports the stopping boundaries for each stage. These are found using the equation by Wang-Tsiatis~\cite{Wang1987} (Section~\ref{sec:MOMS_design_1}). With the stopping boundaries known, the investigator will end the trial only if $m$ upper boundaries or $K-m+1$ lower boundaries are crossed simultaneously.

The stopping boundaries for the multi-outcome DtL design can be found by inverting the two-stage case of Jennison and Turnbull's equation for CP~\cite{jennturn}, giving

\begin{align*}
    f_k & = \frac{1}{\sqrt{\mathcal{I}_{1k}}} \left[ \sqrt{\mathcal{I}_{2k} - \mathcal{I}_{1k}} \; \Phi^{-1}(CP_L) + Z_{\alpha}\sqrt{\mathcal{I}_{2k}} - (\mathcal{I}_{2k} - \mathcal{I}_{1k}) \delta_{1k} \right]\\
    e_k & = \frac{1}{\sqrt{\mathcal{I}_{1k}}} \left[ \sqrt{\mathcal{I}_{2k} - \mathcal{I}_{1k}} \; \Phi^{-1}(CP_U) + Z_{\alpha}\sqrt{\mathcal{I}_{2k}} - (\mathcal{I}_{2k} - \mathcal{I}_{1k}) \delta_{1k} \right],
\end{align*}

where $\mathcal{I}_{1k}, \mathcal{I}_{2k}$ are the outcome-specific information $\mathcal{I}$ for each stage. As above, the investigator stops the trial at the interim if $m$ upper boundaries or $K-m+1$ lower boundaries are crossed simultaneously. Otherwise, some outcomes are dropped at the interim and the trial continues. However, the number of outcomes retained is min$\left( K_{max}, \; \sum\limits_{k=1}^K \mathbb{I} \left(CP_k(\delta_{1k}) > CP_L \right) \right)$. Consequently, if the number of outcomes to be dropped is greater than the number of outcomes that are lower than the lower boundary, the investigator must be able to know which outcomes to be dropped. That is, they must know the CP of the outcomes at the interim. This may be undertaken by, for example, generating a lookup table as part of the design realisation output.

\subsection{Future work}\label{sec:future_work}
There is scope for future work regarding the proposed multi-outcome designs. This may involve generalising the number of stages in the DtL design, which could result in further savings in ESS and ENM. The proposed approaches are for a single-arm trial, while Sozu et al. use a two-arm design~\cite{Sozu}. Extending these designs to two arms would give investigators more options with regards to trial design. Other possible generalisations include allowing interim boundaries and CP boundaries to differ across outcomes for the multi-outcome and DtL approaches respectively and allowing final boundaries to differ, for both approaches. Such lack of generalisation may be considered limitations of the proposed approaches. For the DtL design, it would be worthwhile to undertake a sensitivity analysis to fully explore the effects of varying the interim CP bounds.

It may be possible to divide outcomes into those that must show promise, effectively a subset containing multiple co-primary outcomes, and those among which only a subset are required to show promise. This would be of use if, for example, a treatment is required to show an effect on some safety outcome and simultaneously show an effect on some specified number of efficacy outcomes. Other possible extensions include the introduction of alpha spending, rather than using an overall type-I error-rate, and extending to other types of outcome, for example binary outcomes.

\subsection{Conclusion}
The proposed approaches allow investigators to measure, at least initially, a range of outcomes while reducing the high costs that may be associated with such trials. Furthermore, these approaches offer novel flexibility in the area of multiple-outcome clinical trials, allowing investigators to specify any number of outcomes for which promise must be shown. This is a generalisation of existing designs, which are special cases in comparison as they require promise to be shown on either all outcomes or at least one outcome.

\bibliographystyle{unsrt} 
\cleardoublepage
\bibliography{references} 

\end{document}